\documentclass[12pt,preprint]{aastex}

\pdfoutput=1

\renewcommand {\deg}   {\mbox{$^\circ$}}

\newcommand   {\kms}   {\mbox{km\,s$^{-1}$}}
\renewcommand {\ga}    {\mbox{\rlap{\hbox{\lower5pt\hbox{$\sim$}}}\hbox{$>$}}}
\renewcommand {\la}    {\mbox{\rlap{\hbox{\lower5pt\hbox{$\sim$}}}\hbox{$<$}}}

\begin{document}



\def\kms {\hbox{km{\hskip0.1em}s$^{-1}$}} 
\def\gos #1 {\left({G\over 10^3 G_0 }\right)^{#1}}
\def\ee #1 {\times 10^{#1}}          
\def\msol{\hbox{$\hbox{M}_\odot$}}
\def\lsol{\hbox{$\hbox{L}_\odot$}}
\def\kms{km s$^{-1}$}
\def\Blos{B$_{\rm los}$}
\def\etal   {{\it et al. }}                     
\def\psec           {$.\negthinspace^{s}$}
\def\pasec          {$.\negthinspace^{\prime\prime}$}
\def\pdeg           {$.\kern-.25em ^{^\circ}$}
\def\degree{\ifmmode{^\circ} \else{$^\circ$}\fi}
\def\ee #1 {\times 10^{#1}}          
\def\ut #1 #2 { \, \textrm{#1}^{#2}} 
\def\u #1 { \, \textrm{#1}}          
\def\nH {n_\mathrm{H}}
\def\ddeg   {\hbox{$.\!\!^\circ$}}              
\def\deg    {$^{\circ}$}                        
\def\le     {$\leq$}                            
\def\sec    {$^{\rm s}$}                        
\def\msol   {\hbox{$M_\odot$}}                  
\def\i      {\hbox{\it I}}                      
\def\v      {\hbox{\it V}}                      
\def\dasec  {\hbox{$.\!\!^{\prime\prime}$}}     
\def\asec   {$^{\prime\prime}$}                 
\def\dasec  {\hbox{$.\!\!^{\prime\prime}$}}     
\def\dsec   {\hbox{$.\!\!^{\rm s}$}}            
\def\min    {$^{\rm m}$}                        
\def\hour   {$^{\rm h}$}                        
\def\amin   {$^{\prime}$}                       
\def\lsol{\, \hbox{$\hbox{L}_\odot$}}
\def\sec    {$^{\rm s}$}                        
\def\etal   {{\it et al. }}                     
\def\xbar   {\hbox{$\overline{\rm x}$}}         


\title{Widespread  Methanol Emission\\
from  the Galactic Center}

\author{F. Yusef-Zadeh$^1$,  W. Cotton$^2$, S. Viti$^3$, M.  Wardle$^4$ and M. Royster$^1$}

\affil{$^1$Department of Physics and Astronomy, Northwestern University, Evanston, IL 60208}
\affil{$^2$National Radio Astronomy Observatory,  Charlottesville, VA 22903}
\affil{$^3$Department of Physics and Astronomy, University College London, Gower St. London, WCIE 6BT, UK}
\affil{$^4$Department of Physics \& Astronomy, Macquarie University, Sydney NSW 2109, Australia}


\begin{abstract} 
We report the discovery of a widespread  population of collisionally 
excited methanol J = 4$_{-1}$ to 3$_0$ E sources at 36.2 GHz from the inner 
$66'\times18'$ (160$\times43$ pc) of the Galactic center.  This 
spectral feature was imaged with a spectral resolution of ~16.6 \kms\,  taken from 41 
channels of a VLA continuum survey of the Galactic center region. The revelation of 356 
methanol sources, most of which are  maser candidates,  suggests a large abundance of 
methanol in the gas phase in the Galactic center region. 
There is also spatial and kinematic correlation between 
SiO (2--1) and CH$_3$OH emission from  four Galactic center clouds: 
the +50 and  +20 \kms\, clouds and  G0.13-0.13 and G0.25+0.01.  
The enhanced abundance of methanol 
is accounted for in terms of   induced photodesorption by cosmic rays 
as they travel through a molecular core, collide, dissociate, ionize, and 
excite Lyman Werner transitions of H$_2$.  
A time-dependent chemical model in which cosmic rays drive the
chemistry of the gas predicts CH$_3$OH abundance of $10^{-8}$ to $10^{-7}$ 
on a chemical time scale of $5\times10^4$  to $5\times10^5$ years. 
The average methanol abundance produced by the  release of 
methanol from grain surfaces 
is  consistent with the available data. 
\end{abstract}




\section{Introduction}




Large scale molecular line surveys of the inner few hundred pc of the Galaxy, also known as the 
central molecular zone (CMZ),  have detected emission lines from a wide array of molecular species 
formed in gas phase (e.g., H$_3^+$, CO, CS, HCO$^+$, N$_2$H$^+$) and grain-surface (e.g., SiO, 
CH$_3$OH and NH$_3$) chemistry (Martin-Pintado et al. 1997; Martin et al. 2004; Oka et al. 2005; 
Riquelme, et al. 2010; Jones et al. 2012).  What is interesting about these results is that the 
gas characteristics are similar to those of hot cores associated with protostars and yet the 
distribution of molecular gas is widespread over a few hundred parsecs with a few isolated pockets 
of star formation such as Sgr B2. 
The gas in the Galactic center has 
a higher temperature than the dust temperature $\leq$30K 
(e.g., Fig. 3 of Molinari et al. 2011). This 
characteristic of the warm gas is hard to explain if the hot cores are heated by high mass 
protostars throughout the population of Galactic center molecular clouds.  To investigate the 
chemistry of the gas in the CMZ, we carried out a high resolution survey of methanol emission 
from the inner 66$'\times18'$ of the Galactic center.


Here we report  discovery of a large population of collisionally excited methanol J = 4$_{-1}$ to 3$_0$ E 
emission at 36.2 GHz.
Class I methanol masers are collisionally pumped and 
are generally correlated with outflows in star forming sites (e.g., Voronkov et al. 2006). 
Given  the pervasive distribution of detected maser candidates, 
we suggest instead that enhanced abundance of  methanol
is produced globally by the 
interaction between  
enhanced  cosmic rays  and molecular gas in the CMZ.

\section {Observations and Data Reduction}


The maser observations were conducted with the 
the K. Jansky Very Large Array (VLA) of the National Radio Astronomy Observatory\footnote{The
National Radio Astronomy Observatory is a facility of the National Science Foundation, operated
under a cooperative agreement by Associated Universities, Inc.} (NRAO)
in its C configuration as part of a continuum survey in which one subband was centered
on the rest frequency of the rotational class I methanol 
masers (36.169265
GHz). This subband had 64 $\times$ 2 MHz channels giving a velocity resolution of $\approx$ 16.6
\kms. Each of the $\ge$900 single snapshot pointings were imaged independently and the results combined
in a linear mosaic.
These observations were made in six sessions between 2012 February 21 and
2012 March 10 under project code  12-120A. 
J1744-3116 was used as the phase calibrator and 
3C286 was used as the photometric and bandpass calibrator with an 
assumed flux of 1.70 Jy and an uncertainty of 5\%. 
Using  the Obit package
\footnote{http://www.cv.nrao.edu/bcotton/$\sim$Obit.html}, 
calibration,  editing and imaging were  done on each session independently and
consisted of  determining first instrumental group delay offsets  from
observations of 3C286 and J1744-3116,  and applying  to all data.
The methanol features were much narrower in frequency than a channel,  so
the stronger features gave significant Gibbs ringing in frequency 
due
to the limited range of lags used by the correlator.
The resolution is approximately 1.8'' $\times$ 0.7'' with RMS noise 
$\approx$ 2.5 mJy/beam.
The image cubes for each pointing 
were collapsed to a single plane containing only
``significant'' pixels and all pixels within 8 cells of significant
pixels (see Fig. 2d).
These were the brightest points in each image plane which was
in excess of 6 times the off--source RMS and more than 0.15 times the
brightest pixel in the channel image.
Maser candidates  were selected from the combined, collapsed cubes and
elliptical Gaussians fitted.
Note, this procedure will select continuum as well as maser candidates.
Maser candidates  were then tested using the spectrum extracted at the
appropriate location from the combined spectral cube (see Fig. 2d).
Spectral features were accepted if the channel with the highest flux
density was in excess of 5 times the RMS of the other channels.




\section{Results}

Figure 1a reveals  a color coded distribution of 356 methanol sources found 
in the region between Galactic longitude $-30' < l < 36'$ and latitude $-12' < b < 6'$.  The size 
of each methanol source is proportional to the line flux of each source. 
 The list of methanol sources of  Table 1 in 10 columns 
gives:  the source number, Galactic 
coordinates, the spatially integrated line flux, the center velocity, 
the lower limit to the brightness temperature (T$_{\rm b}$) estimated 
from the maximum value in the spectrum interpolated at the
fitted centroid of the spot in the collapsed image, 
the  fitted values of the angular size and references. 
Given the large channel width (16.6 \kms),  
we can only identify maser candidates  by
their high estimated 
T$_{\rm b}$, without information on the linewidths.
For unresolved point sources, the   line flux ranges   between  0.19 and 468.31  Jy \kms,    corresponding to
brightness temperatures  
ranging   between   T$_{\rm b}$=8.5K and 
2.1$\times10^4$K,  respectively. 
Typical masers have linewidths that are one to two orders of magnitudes 
smaller than the channel width we used, thus these sources could be much 
brighter if measured with smaller channel widths.   
As for sources with low brightness temperature, 
it is possible that these weak sources trace thermal methanol gas. However, thermal 
emission is generally spatially extended and the present snapshot survey 
does not have the sensitivity to detect 
extended emission on  a scale  $\ge22''$. 
To test that bright sources in Table 1 include previously detected masers, 
we compare
our sources with those of Sjouwerman et al. (2010; hereafter SPF)
who had spectroscopically identified 
36.2 GHz masers in the 50 \kms\, molecular cloud. 
SPF tabulates  ten methanol masers 
(number 10, 6, 7, 8, 9 and  4 in   their Table 1) 
ranging  in brightness temperature 6.6$\times10^4 < \rm T_b < 2.3\times10^6K$.  
Six of the brightest  sources coincide with 
sources that are detected in our survey  (number  160, 164, 165, 167, 168, 169 in our Table 1) 
having  similar LSR velocities. 
These spatial  and spectral coincidences  
confirm   maser candidate  identification  of bright  sources  in our survey.  

Figure 1b shows the distribution of all detected methanol sources with their 
peak velocities superimposed on a grayscale continuum image 
of the Galactic center at 5 GHz. 
The peak velocities are accurate to within 10 \kms\, and can not 
identify multiple  velocity components. 
Prominent radio continuum sources Sgr A, the nonthermal 
vertical and thermal arched filaments of the radio Arc, Sgr B1 and Sgr AC (the region between Sgr A and Sgr 
C) are labeled.  The largest concentration of maser candidates lies 
in Sgr AC near $l\sim40'$ and the radio Arc near 
$l\sim0.1^{\circ}-0.2^{\circ}$. 
We note a total of 200 and 150 sources distributed between $-0.5^{\circ} < l 
< 0^{\circ}$ and $0^{\circ} < l < 0.5^{\circ}$, respectively, suggesting an asymmetry with respect to 
Galactic longitude. 
We divide the source list into strong or  weak  maser candidate  sources  
with line fluxes  greater than or less than 10 Jy \kms\ ($T_b\sim$446K for unresolved sources), 
respectively.  This threshold is selected because the gas temperature 
in the CMZ is much less than 446K. 
Figure 1c shows the distribution of all  methanol maser candidates superimposed on a 
grayscale 24$\mu$m image showing similar asymmetry in terms of the number of YSO candidates 
(Yusef-Zadeh et al. 2009). This implies 
that star formation processes  may be contributing to the origin of this asymmetry.


To examine the  relationship between CH$_3$OH (36.2 GHz) maser candidates 
and dust emission,  
Figure 2a and 2b   show the distribution of weak and strong   methanol  
emission   superimposed on 
the distribution of dust clouds 
at 850$\mu$m,  respectively (Pierce-Price et a. 2000). 
Methanol sources  generally follow 
 the ``bow-tie'' dust layer which  runs parallel to  the Galactic plane. 
The brightest methanol sources in Figure 2b 
show a good   correlation of maser candidates  and dust emission from the CMZ. 
To examine the central region of the CMZ in more detail, 
Figure 2c views  the distribution of weak maser  candidates  
superimposed on a grayscale  image of 
SiO (2-1) line emission map (Tsuboi et al. 2011). 
Figure 2d shows the  spectrum of a representative maser candidate and 
a collapsed  image showing  the distribution of bright sources in G0.13-0.13. 
A correlation between methanol 
maser candidates  and SiO line emission suggests 
that the chemistry in producing enhanced  methanol and SiO line emission is similar.  
This is interesting because the abundance of SiO produced in the ambient gas is too low, 
thus grain surface chemistry is needed to enhance the abundance of 
both SiO and CH$_3$OH throughout the Galactic center. 
In the following, we  briefly discuss 
the distribution of methanol sources   toward  four  
Galactic center molecular  clouds. 


\noindent{\bf The 50 \kms\, Cloud:} 
The 50 \kms\, M--0.02--0.07, is physically interacting with an 
expanding shell of the Sgr A East SNR G0.0+0.0 (e.g., Tsuboi et al. 2011). The presence of OH (1720 MHz) 
masers at the site of the interaction suggests that the abundance of OH 
is enhanced behind a supernova shock 
driving into the molecular cloud (Yusef-Zadeh et al. 1999; Wardle 1999).  Figure 3a shows contours of SiO 
(2-1) line emission superimposed on a 5 GHz  continuum image. The crosses represent the positions of 18 CH$_3$OH 
(36.2 GHz) maser candidates
 detected toward this cloud.  The peak velocities,  which 
are drawn next to the position of 
individual 
maser candidates, range between 20 and  50 \kms\, with the exception of one source showing 
a peak velocity of --133 \kms.  
The largest concentration of maser candidates  coincide with a region where 
SiO (2-1) emission is strong but are offset by $\sim30''$ (1.2 pc) from the compact HII regions.
The circumnuclear ring orbiting Sgr A* lies to the west of the 50 \kms\, cloud in Figure 3a   
and shows a lack of  CH$_3$OH (36.2 GHz) maser candidates (see also SPF 2010). This result  is consistent with 
CH$_3$OH (96 GHz) observations of the inner 10$'$ of the Galactic center by 
Stankovi\'c et al. (2007),  who suggest that  
strong UV radiation  from young stellar clusters at the Galactic center is 
responsible for   the destruction of methanol  in the 
circumnuclear ring.  



\noindent {\bf The 20 \kms\, Cloud}
Another prominent Galactic center molecular cloud 
within four arcminutes of Sgr A*  is 
the 20 \kms\,  molecular cloud  M-0.13-0.08.  
Figure 3b  shows contours of SiO (2-1) line emission 
from this cloud 
superimposed on a grayscale 5 GHz continuum emission. 
Maser Candidates  are mainly concentrated in the region where SiO (2-1) 
line emission peaks having velocities $\sim$20  and $\sim$0 \kms\, to the north 
and south of the cloud, respectively. 
The distribution of SiO (2-1), CH$_3$OH (36.2 GHz) and CH$_3$OH (96 GHz; Stankovic et al. 2007) 
are  remarkably similar to each other suggesting that 
the abundance of SiO and CH$_3$OH in the gas phase are  enhanced. 
A  circular-shaped compact  HII region Sgr A-G lies to the north 
and two extended nonthermal filaments  Sgr A-E (G359.88-0.08) and Sgr A-F (G359.90-0.06)
lie to the south.  
Both nonthermal filaments have X-ray counterparts showing a nonthermal spectrum 
(Sakano et al. 2003; Yusef-Zadeh et al. 2005), perhaps due to  upscattering
of far-infrared photons from dust emission of  the 20 \kms\, cloud 
by  the relativistic electrons in the radio filaments.



\noindent {\bf G0.13-0.13}
This cloud lies along the
nonthermal filaments of the radio Arc near l$\sim0.2^{\circ}$, the most
prominent network of magnetized filaments in the Galactic center.
The kinematics of CS line emission from G0.13--0.13                 
suggests an expansion of molecular gas into the nonthermal filaments             
(Tsuboi et al.  1997).
Figure 3c shows the distribution of CH$_3$OH (36.2 GHz) emission from this cloud. 
Like the 50  and 20 \kms\,  molecular clouds, 
45 maser candidates    with velocities that 
range  between 0 and 50 \kms\, 
appear to trace the distribution of SiO (2-1) line emission with similar velocity. 
We also find  three  high velocity maser candidates  at the eastern and western boundaries of the 
cloud. This cloud has recently been studied in detail showing compelling evidence 
for the interaction of molecular gas with nonthermal electrons (Yusef-Zadeh et al. 2012).  

\noindent {\bf G0.25+0.01}
G0.25+0.01 is a quiescent giant molecular cloud that coincides with the 
 darkest cloud at mid-IR wavelengths in the so-called ``Dust Ridge" (Lis \& Carlstrom 1994).
This cloud has a mass 
of 1.4$\times10^5$ \msol\, and 
exhibits little    
star formation (Lis and Carlstrom 1994; Immer et al.  2012).
Figure 3d shows the spatial distribution of 
CH$_3$OH (36.2 GHz) maser candidates
and are 
represented  as crosses with corresponding velocities
superimposed on a 24$\mu$m image.
Contours of SiO (2-1) line emission are also superimposed on 
Figure 3c.  
Like the above discussed clouds, maser candidates
 in G0.25+0.01 
follow the distribution of SiO (2-1)  and dust emission.  
There are no signatures of 
 high mass star formation in this cloud,  such as  
compact HII regions excited by OB stars,  CH$_3$OH (6.6 GHz) masers,   or 
shocked molecular outflows associated with  protostars. 


\section{Discussion}

The distribution of methanol emission throughout the Galactic center raises an important
question: what is the mechanism by which methanol molecules are released off dust grains to
enhance its abundance in the gas phase over such a widespread region. 
The 
tight correlation between SiO and CH$_3$OH emission suggests that grain surface 
chemistry is responsible for their production. Methanol is formed by hydrogenation of 
CO on interstellar grains at temperatures of 10-20K (Watanabe and Kouchi 2002) and is 
released into the gas phase by 
heating provided by UV radiation, by shocks due to cloud-cloud collisions, and by protostellar
outflows. 
These processes are known to be 
important in star forming regions where hot molecular cores are formed and in 
cloud-cloud collisions where shocks drive into clouds. UV radiation can be important to 
release methanol but it is not possible to produce widespread methanol emission from 
dense, self-shielded Galactic center molecular clouds.  The difficulty with large-scale 
shocks produced by cloud-cloud collisions is that methanol maser candidates
are seen deep within 
dense giant molecular clouds and these shocks are most effective at the surface of 
clouds where the interaction takes place.  Shocks can be generated locally in star 
forming sites. However, with the exception of Sgr B2, there is no evidence for 
widespread on-going star formation throughout the CMZ. The four giant  molecular clouds 
that were discussed previously are examples in which on-going massive star formation 
has low efficiency (e.g., Immer et al. 2012).   A more detailed discussion of 
additional issues related to 
 enhanced SiO abundance  in the Galactic center is discussed elsewhere (Yusef-Zadeh et al. 
2013).  



Instead,  enhanced cosmic rays in the Galactic center interacting with molecular clouds 
provide a mechanism for  globally enhancing  the abundance of methanol. Recent large-scale studies 
indicate that the cosmic ray ionization rate over the inner few hundred pc is $10^{-15}$ to 
10$^{-14}$ s$^{-1}$,  which is about one-to-two orders of magnitude larger than in the Galactic 
disk (Yusef-Zadeh et al. 2012; Ao et al. 2012).
Induced photodesorption by cosmic rays traverse a molecular
cloud and collide with, dissociate, ionize and excite Lyman and Werner
transitions of H$_2$ (Prasad and 
Tarafdar 1983; Roberts et al. 2007). The FUV emission resulting from these interactions can heat 
dust grains and evaporate methanol. There are other mechanisms that can evaporate methanol but 
total desorption of icy mantles, in a shielded         
environment, could only be explained by high rates of cosmic ray ionization (Roberts et al.
2007). 



LVG modeling of CH$_3$OH emission from the inner 30pc of the Galactic center at 96 and 
242 GHz requires a two-component model (Stankovi\'c, et al. 2007). 
The H$_2$ density and temperature are  n$=10^4$ cm$^{-3}$ and 
T$\sim$90K in the warm phase with  column density N$_{CH_3OH}\sim2.6\times10^{15}$ 
cm$^{-2}$, whereas in the cold phase, N$_{CH3OH}\sim8\times10^{15}$ cm$^{-2}$ is 
estimated for n$=5.5\times10^6$ cm$^{-3}$, T$=15$K (Stankovi\'c, et al. 2007).  
Typical column densities  of H$_2$  toward Galactic center clouds is $10^{23}$ to $10^{24}$ cm$^{-2}$, 
thus the 
 abundance of methanol ranges roughly  between  
$10^{-7}$ to $10^{-9}$ in these two phases. 
To explore the effect of high 
cosmic-ray ionization rates on the chemistry of the gas, 
we use a gas-grain time dependent 
chemical model,  UCL\_CHEM, 
to investigate the abundance of different species and compare them 
with observed values (Viti et al. 2004). 
This  model initially follows the collapse of a prestellar 
core, and subsequently the warming and evaporation of grain mantles due to 
either the increase of temperature and/or an enhanced cosmic ray ionization rate.  
Given that T$_{\rm dust}\leq$30K, 
thermal evaporation is insignificant in this model.  
Figure 4 shows the abundance of CH$_3$OH, SiO, OH and  NH$_3$,
 as a function of time, for a grid of models. 
These species are either produced on the surface of the grains (CH$_3$OH, 
NH$_3$), or undergo an enhancement due to the release of the parent species from the grain mantles 
(SiO, OH). 
In these models, we  varied the density from 10$^4$ cm$^{-3}$ 
to 10$^6$ cm$^{-3}$  and the cosmic ray ionization rate from 5$\times10^{-16}$  to 5$\times10^{-14}$ 
s$^{-1}$. The models presented in Figure 4 give suitable CH$_3$OH abundances before they 
are destroyed. 
 It turns out that the very process that is responsible to enhance the abundance of methanol in gas phase also destroys it on 
a time scale 10$^4 - 10^5$ years. Figure 4 shows that methanol for low cosmic ray ionization rates lasts longest for both high 
and low density gas. In other words, as the cosmic ray ionization rate decreases, the chemical time scale for destruction of 
methanol increases. Methanol is highly volatile at high cosmic ray ionization rates, thus 
has  a short destruction time scale.  This 
is somewhat puzzling given that methanol emission is detected throughout the CMZ,  which consists of several giant molecular 
clouds distributed within a couple of hundred pc. One possibility to account for this behavior is that methanol is constantly 
being replenished, thus increasing the destruction time scale. The reformation of methanol on the surface of dust grains can 
occur by the cold and dense component of the gas observed in the CMZ (H\"uttmeister et al. 1998) during which methanol is 
being ejected from grain surfaces by cosmic rays. Other hydrogenated species do not get destroyed as fast as 
methanol in this picture. The other possibility is that we may be seeing a relatively short lived phase of the gas. This is 
because the medium is clumpy with several cores at different ages. Both these possibilities will be investigated theoretically 
in more detail elsewhere.



In summary, we carried out a 
survey of the Galactic center which resulted in the discovery of a 
large number of probable methanol masers at 36.2 GHz. These maser candidates
 were detected as part of continuum 
observations at 35 GHz. 
We found  a  strong correlation of maser candidates
and molecular gas distribution in the  
central molecular zone. 
The identification of methanol emission as maser lines needs to be further investigated
because of the poor spectral resolution employed in the continuum settings. 
While maser features are much narrower in frequency than our single channel width, we can put
a lower limit on the brightness temperature of our maser candidates.
It is possible that detected maser candidates  have broad linewidths 
because they arise in molecular clouds  with large linewidths or 
are contaminated by cluster of masers with 
different velocities.  Future  spectral line measurements 
will be able to distinguish  thermal and maser  emission.

\acknowledgments
We thank the referee for useful comments. 
This research is supported in part by grant
AST-0807400 from the NSF the National Science Foundation.



\begin{figure} \center 
\includegraphics[scale=.70,angle=0]{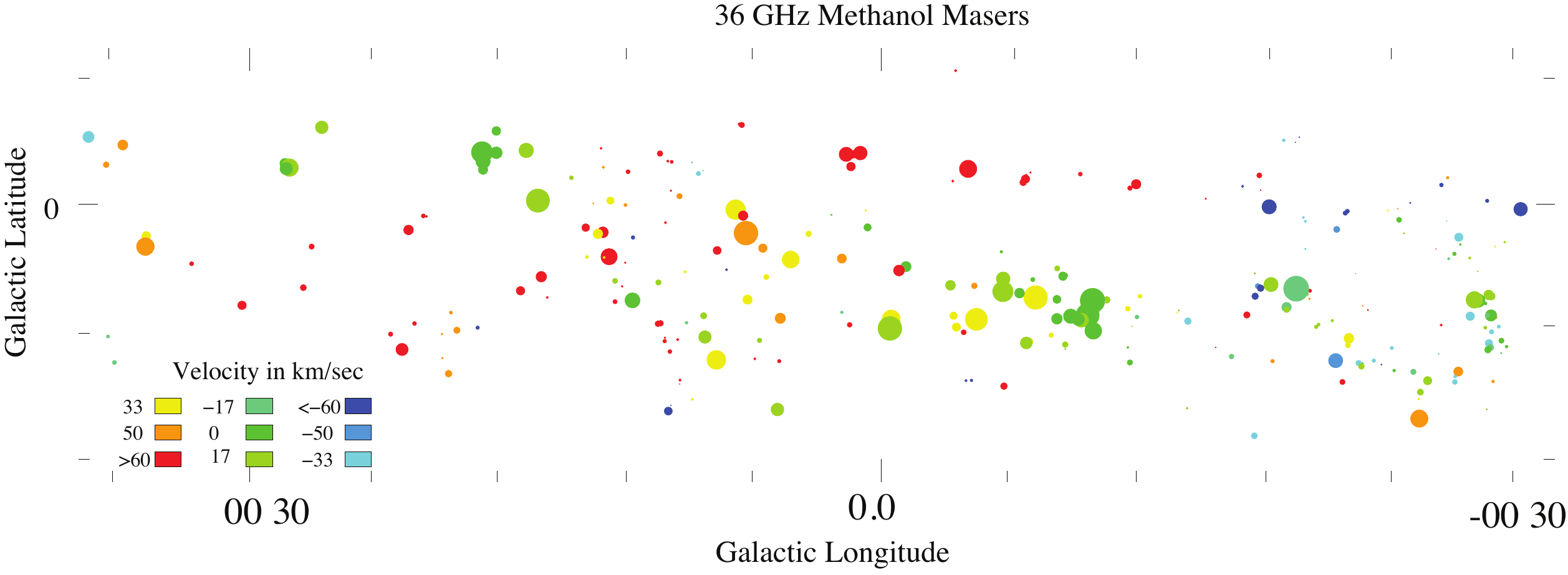} 
\includegraphics[scale=.70,angle=0]{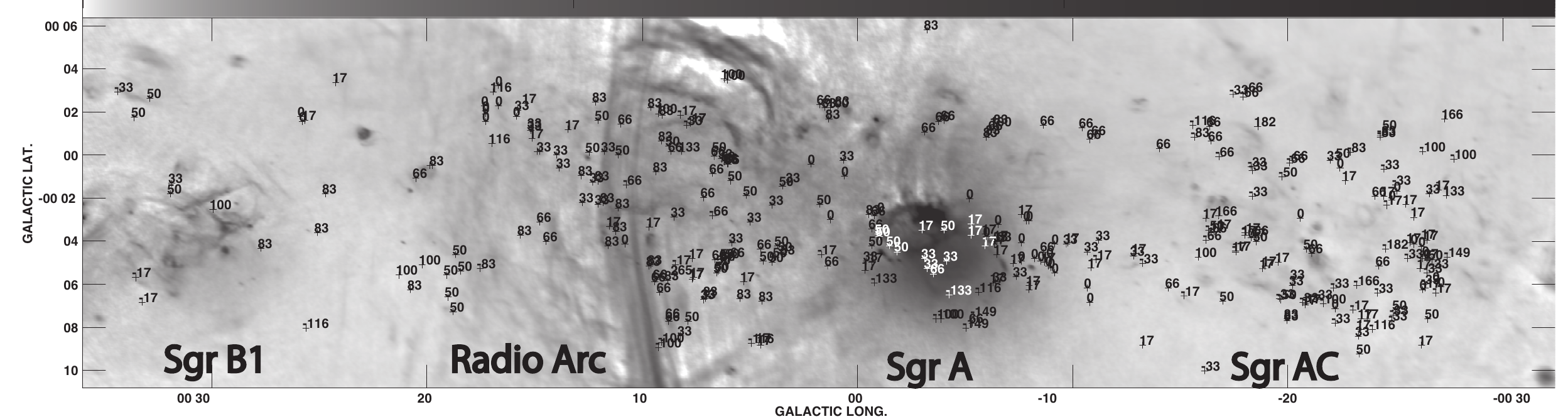} 
\includegraphics[scale=.75,angle=0]{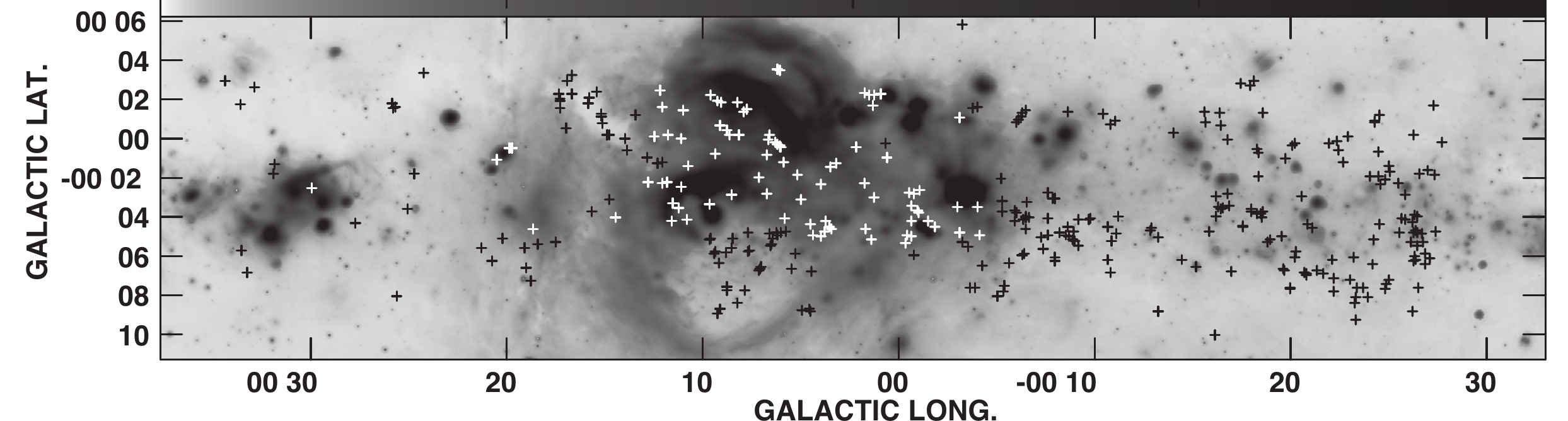} \caption{ {\it 
(a) Top} Maser candidates in (l,b) coordinates. 
Color indicates 
velocity and the size of the symbol is proportional to the brightness. 
The line flux ranges between 0.19 to 468.31 Jy \kms. {\it (b) Middle} 
Methanol sources, with their  velocities 
in \kms, (crosses) superimposed on a grayscale 5 GHz image 
(grayscale flux -0.1 to 300 mJy beam$^{-1}$). 
{\it (c) Bottom} 
Similar to (b) except a 24$\mu$m image (grayscale range 0 to 8; 
Yusef-Zadeh et al. 2009).
}
\end{figure}

\begin{figure}
\center
\includegraphics[scale=0.65,angle=0]{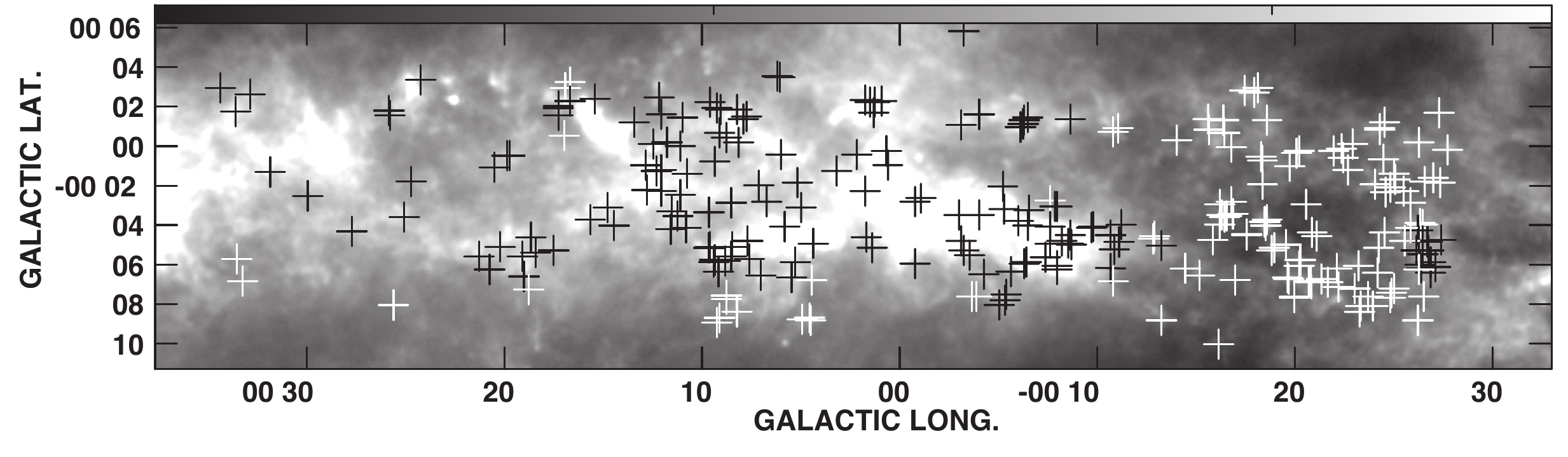}
\includegraphics[scale=0.65,angle=0]{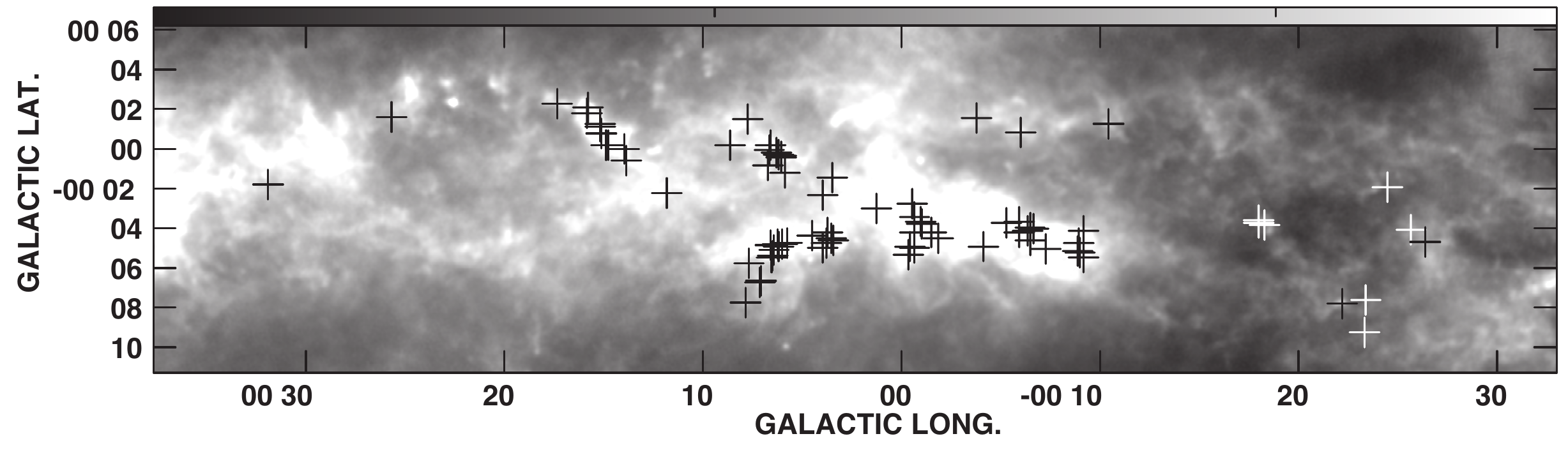}
\center
\includegraphics[scale=0.35,angle=0]{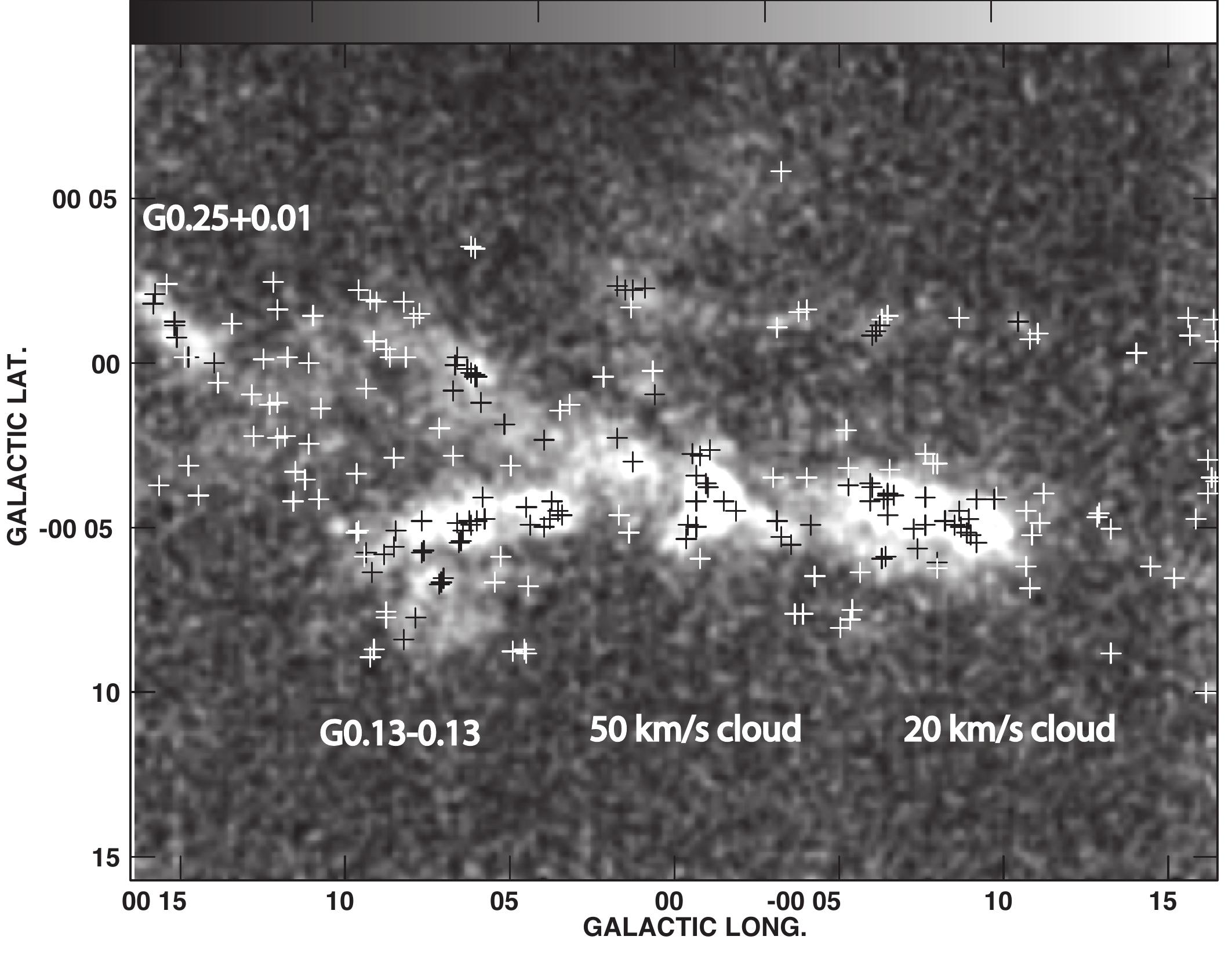}
\includegraphics[scale=0.45,angle=0]{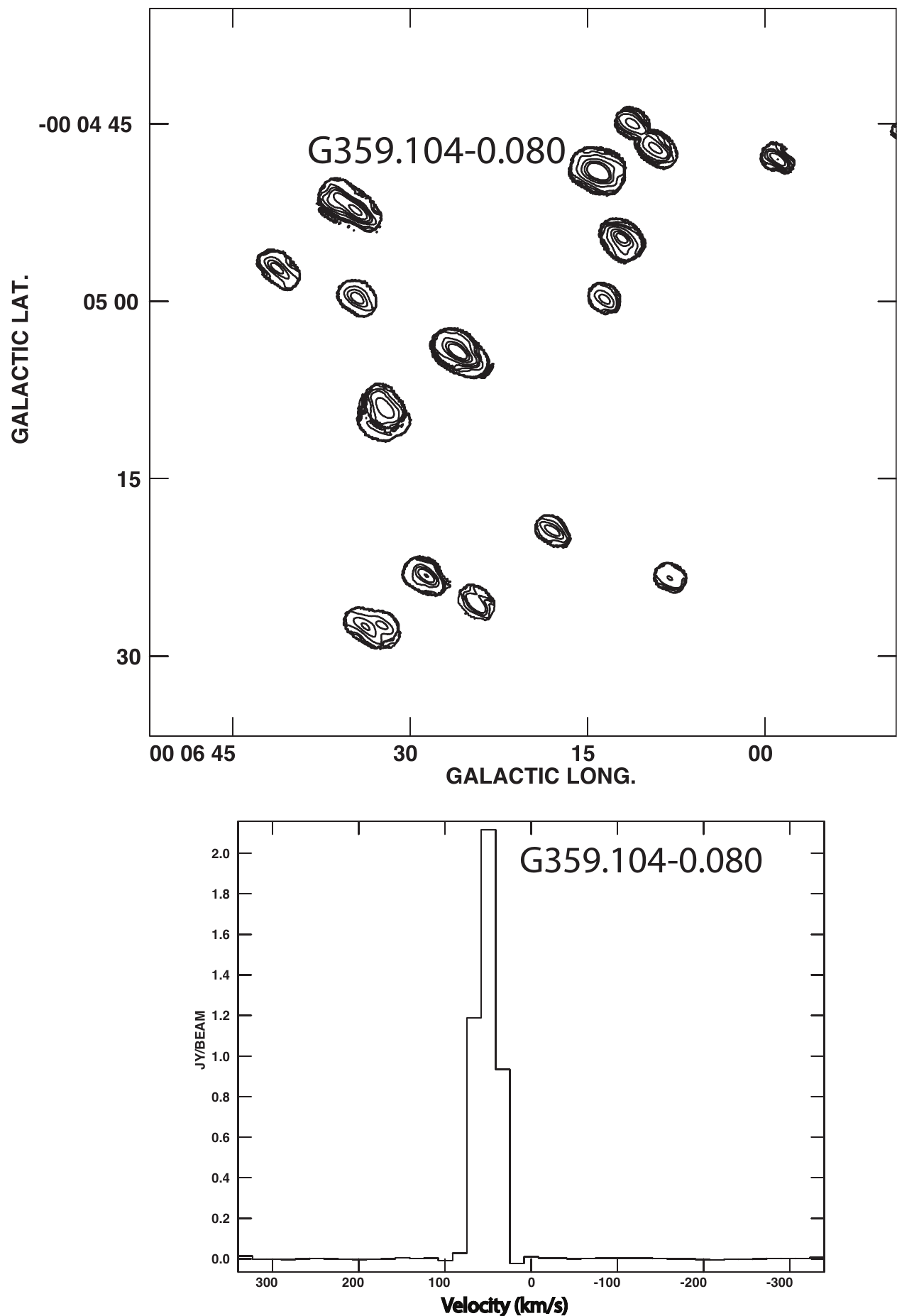}
\caption{
{\it (a) Top}  
Weak  maser candidates (plus signs)
on an 850$\mu$m map  (grayscale range 0 to 5; Pierce-Price et al.  2000).
{\it (b) Middle} 
Identical  to (a) except  bright maser candidates
 are shown. 
{\it (c) Bottom Left}  
All  maser candidates on 
a map of SiO (2-1) line emission 
integrated between  --152 to 197 \kms\, 
(grayscale range 0 to 5; Tsuboi et al. 2011).
{\it (d) Bottom Right} A smoothed sample
spectrum of a bright maser candidate  G359.204-0.080 
(source 113 in Table 1) 
in G0.13-0.13 and a collapsed 36.2 GHz image of G0.13-0.13. 
Contour levels are (0.1, 0.15, 0.2, 0.25, 0.5, 0.75, 1, 1.5, 2.5, 3.5, 4.5)$\times$0.5 
Jy per  fitted beam size  shown in Table 1). 
}
\end{figure}

\begin{figure}
\center
\includegraphics[scale=0.35,angle=0]{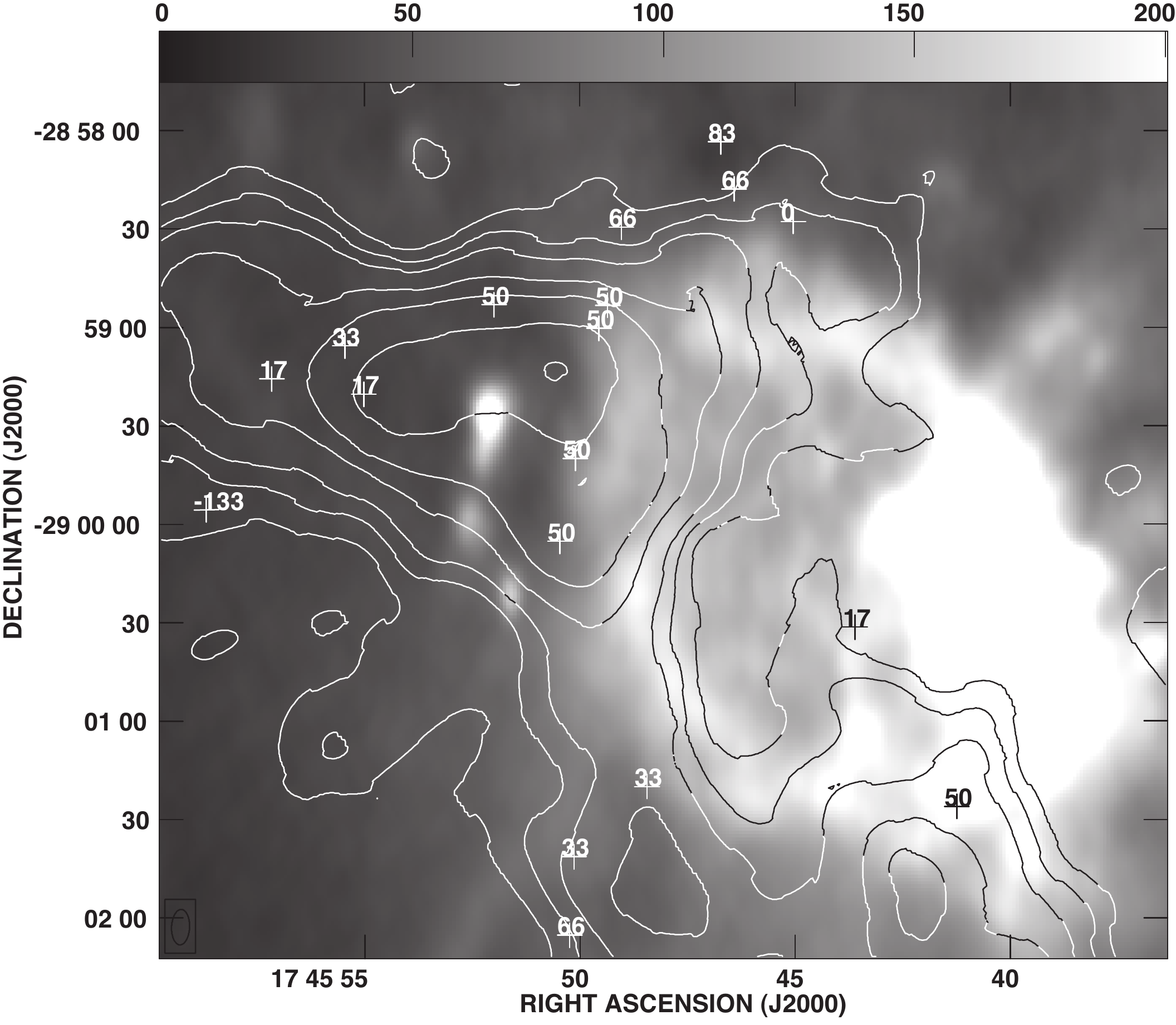}
\includegraphics[scale=0.35,angle=0]{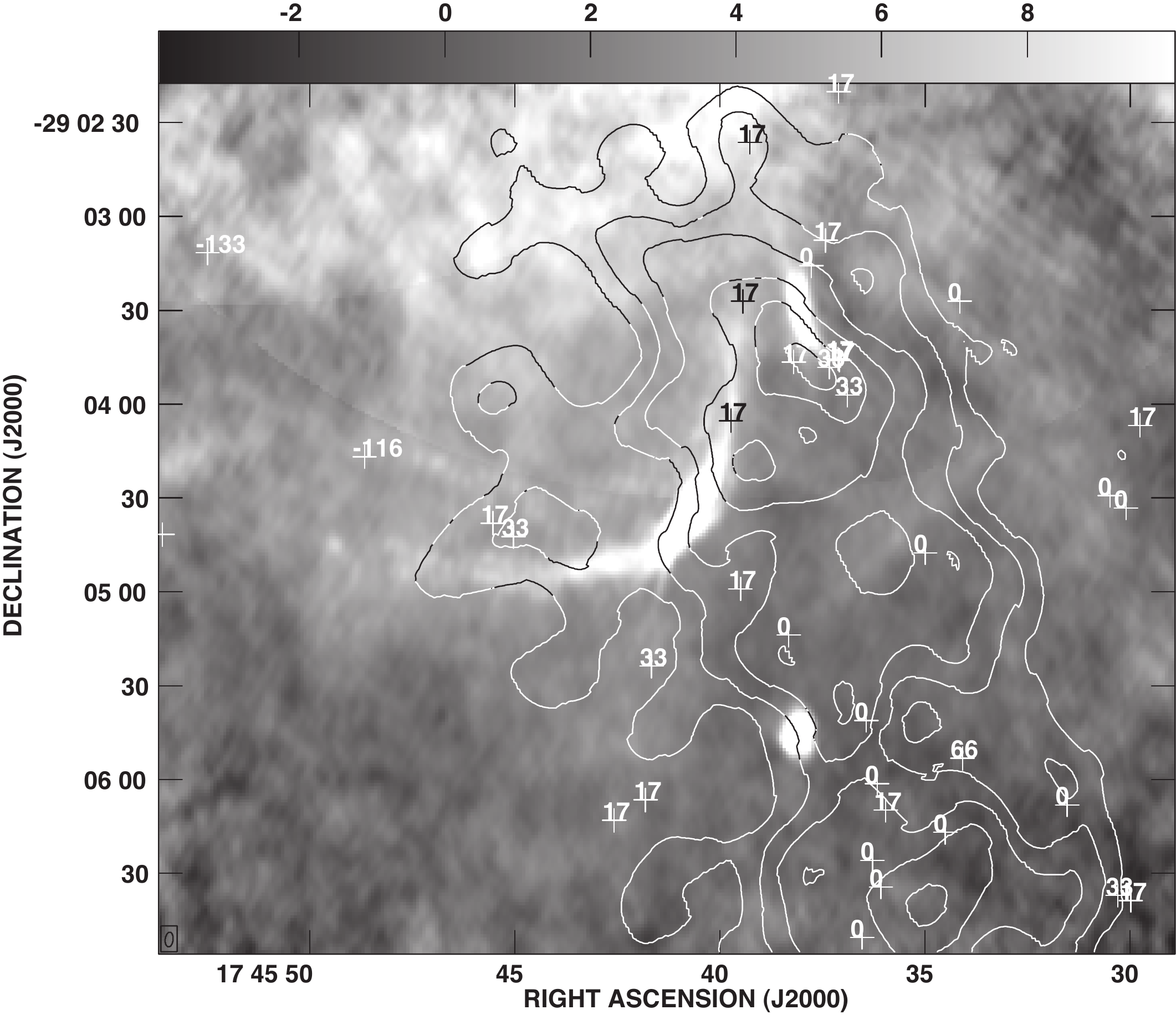}\\
\includegraphics[scale=0.35,angle=0]{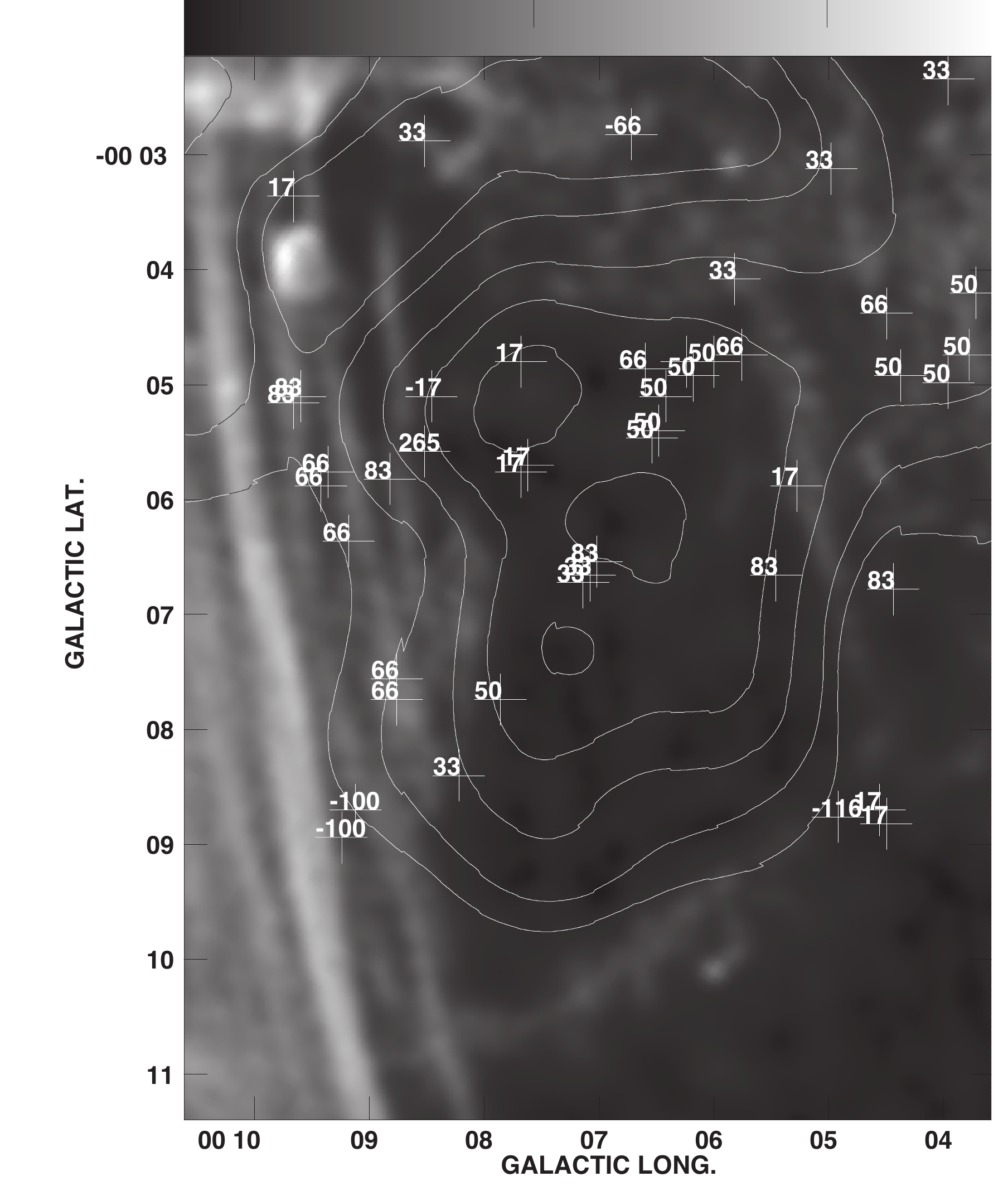}
\includegraphics[scale=0.35,angle=0]{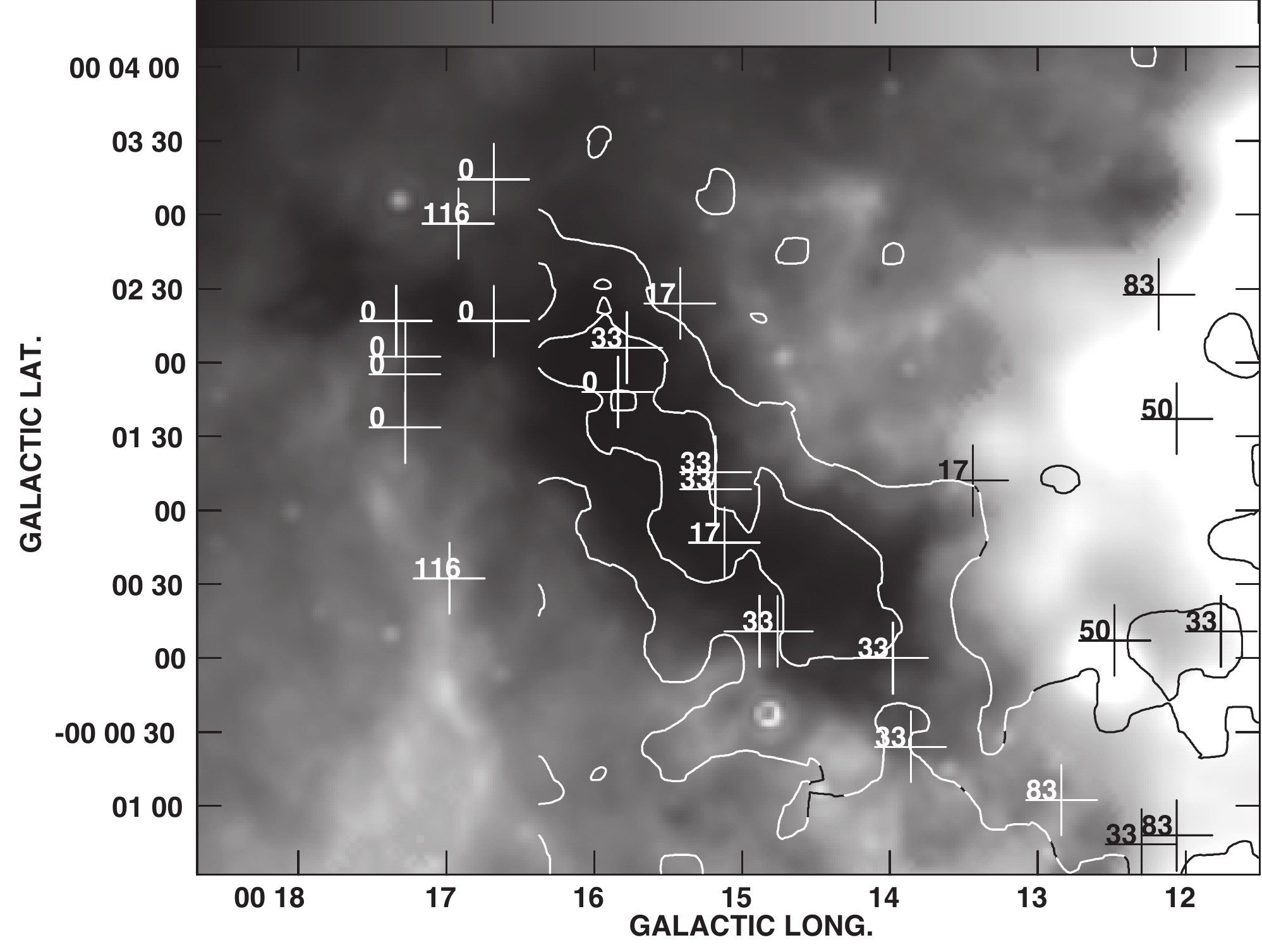}
\caption{
{\it (a) Top Left}  
CH$_3$OH  (36.2 GHz)  maser candidates (crosses) and LSR velocities 
on the shell-type SNR G0.0-0.0 image and its four compact HII regions. 
Contours of SiO (2-1) line emission integrated between  0 and 50 \kms\, 
on the grayscale 5GHz  image. 
{\it (b) Top Right}  
Similar to (a) except the 20 \kms\, molecular cloud. 
{\it (c) Bottom Left}  
Contours of  CS (1-0) line emission integrated  between 0 and 50 \kms\,
from G0.13--0.13  with a
resolution of 45'' (Tsuboi et al. 1997) on 
a 1.4 GHz   image (grayscale range -9.4 to  128) 
Contour levels are 2, 4, 6, 8 and 10 \kms\,  K (T$^*_A$).
{\it (d) Bottom Right}
CH$_3$OH  (36.2 GHz) maser candidates  and contours of SiO (2--1) line emission 
on a  24$\mu$m  map of  
G0.25+0.01 (grayscale  range 0.11 to 1.5).  
}
\end{figure}

\begin{figure}
\center
\includegraphics[scale=0.7,angle=0]{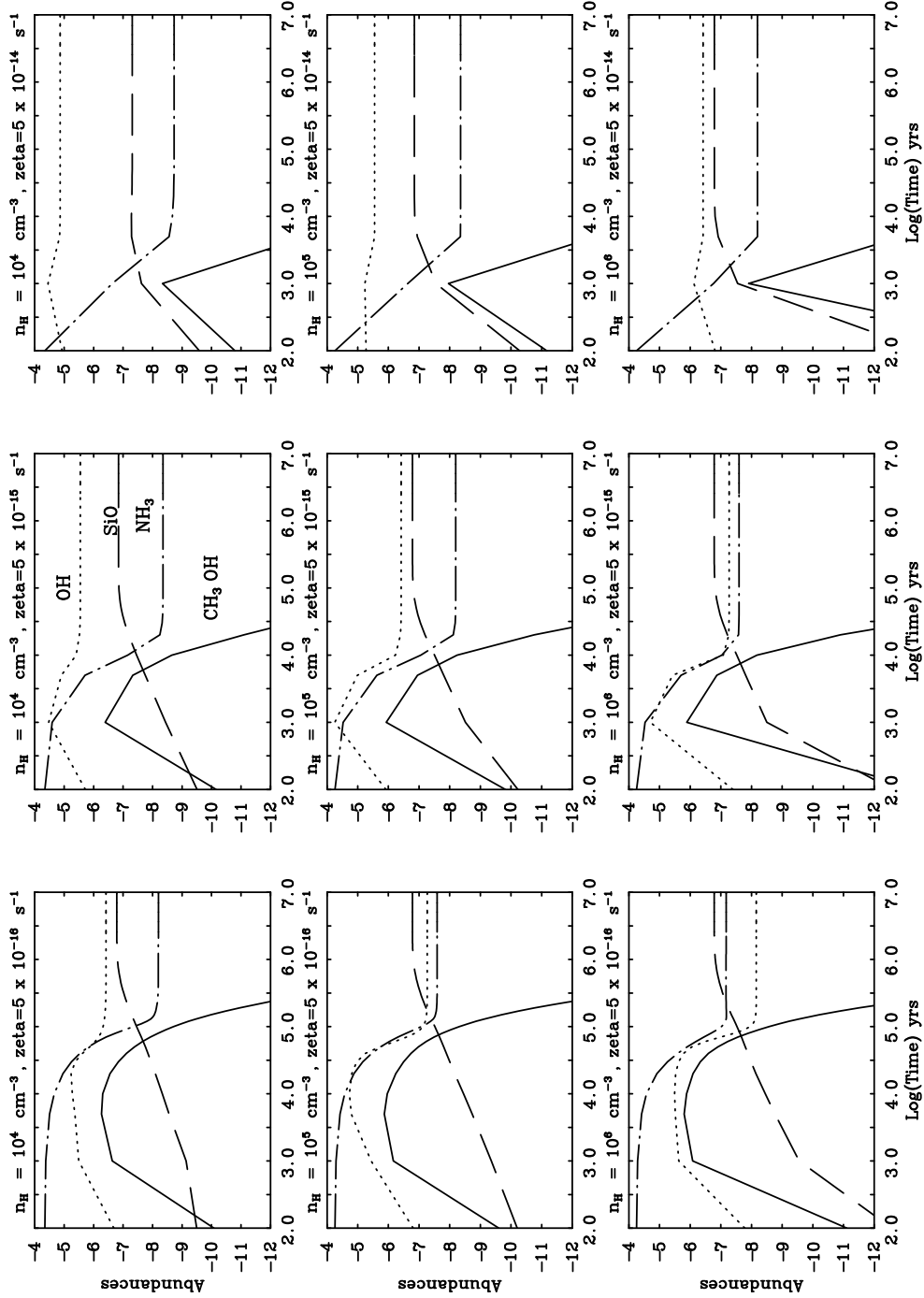}
\caption{
The abundance of CH$_3$OH, OH, NH$_3$ and  SiO as a function of time in years for different
values of cosmic ray ionization rate and molecular gas density
using  the UCL\_CHEM  time-dependent gas-grain chemical model (Viti et al. 2004).
}
\end{figure}

\begin{deluxetable}{ccccccccccc}
\label{Catalog}
\tablecaption{Detected CH$_3$OH Sources in the Survey}
\tabletypesize{\scriptsize}
\tablewidth{0pt}
\tablehead{
\colhead{Source} & \colhead{G. long} & \colhead{G. lat} & \colhead{Flux} & \colhead{Vel} & \colhead{T$_B$} &
    \colhead{Fit Major} & \colhead{Fit Minor} & \colhead{PA} & \colhead{Ref.}\\ 
      & \colhead{$\circ$} & \colhead{$\circ$} & \colhead{Jy km s$^{-1}$} & \colhead{km s$^{-1}$} & 
    \colhead{$^{\circ}$K $\times$ 10$^{2}$} & \colhead{"} & \colhead{"} & \colhead{$\circ$} & \colhead{}
}
\startdata
1   &   0.57307 &  0.04923 &   7.32 &  -33.2 &   63.14 $\pm$  12.60 & 6.00 & 0.66 &   3.7 &    \\
2   &   0.56038 &  0.02935 &   3.74 &   49.7 &   24.77 $\pm$   4.38 & 6.00 & 0.86 &   2.9 &    \\
3   &   0.55883 & -0.09493 &   2.00 &  -16.6 &   23.00 $\pm$   4.17 & 2.27 & 1.31 &   3.8 &    \\
4   &   0.55429 & -0.11371 &   2.63 &  -16.6 &   29.65 $\pm$   4.94 & 2.24 & 1.35 &   5.2 &    \\
5   &   0.54816 &  0.04345 &   6.57 &   49.7 &   44.01 $\pm$   6.30 & 6.00 & 0.85 &   4.7 &    \\
6   &   0.53183 & -0.02997 &  11.49 &   49.7 &   51.91 $\pm$   6.19 & 6.00 & 1.26 &   3.3 &    \\
7   &   0.53134 & -0.02220 &   5.67 &   33.2 &   51.40 $\pm$   7.58 & 3.73 & 1.01 &   0.2 &    \\
8   &   0.49866 & -0.04223 &   2.68 &   99.5 &   33.00 $\pm$   5.07 & 2.62 & 1.06 &   4.3 &    \\
9   &   0.46208 & -0.07243 &   5.55 &   82.9 &   63.26 $\pm$   5.93 & 2.63 & 1.14 &  -0.4 &    \\
10  &   0.43104 &  0.03018 &   6.49 &    0.0 &   52.39 $\pm$   6.99 & 4.92 & 0.86 &   2.7 &    \\
11  &   0.43031 &  0.02641 &   8.13 &    0.0 &   55.03 $\pm$   6.55 & 5.37 & 0.94 &   1.9 &    \\
12  &   0.42780 &  0.02715 &  11.42 &   16.6 &   89.83 $\pm$   8.74 & 5.05 & 0.86 &   1.9 &    \\
13  &   0.42732 & -0.13414 &   0.35 & -116.0 &   15.41 $\pm$   3.06 & 1.19 & 0.65 & 196.1 &    \\
14  &   0.41767 & -0.05960 &   3.98 &   82.9 &   41.43 $\pm$   6.33 & 3.01 & 1.09 &   5.8 &    \\
15  &   0.41177 & -0.02976 &   3.42 &   82.9 &   26.16 $\pm$   4.88 & 4.56 & 0.98 &-183.6 &    \\
16  &   0.40449 &  0.05625 &   8.29 &   16.6 &   49.28 $\pm$   9.65 & 4.75 & 1.21 &   1.2 &    \\
17  &   0.35465 & -0.09316 &   2.87 &   99.5 &   29.34 $\pm$   5.14 & 2.55 & 1.31 &  -7.2 &    \\
18  &   0.34644 & -0.10415 &   7.97 &   82.9 &   23.40 $\pm$   3.37 & 6.00 & 1.94 &  -0.9 &    \\
19  &   0.34168 & -0.01792 &   6.27 &   66.3 &   57.00 $\pm$   3.18 & 2.87 & 1.31 &-190.3 &    \\
20  &   0.33745 & -0.08545 &   2.50 &   99.5 &   38.92 $\pm$   5.08 & 1.94 & 1.13 & -11.7 &    \\
21  &   0.33109 & -0.00765 &   2.64 &   66.3 &   26.37 $\pm$   3.20 & 3.08 & 1.11 &   0.3 &    \\
22  &   0.32893 & -0.00808 &   1.36 &   82.9 &   15.82 $\pm$   3.13 & 2.90 & 1.01 & -10.8 &    \\
23  &   0.31756 & -0.09318 &   1.13 &   49.7 &   17.90 $\pm$   3.45 & 1.64 & 1.31 &-199.4 &    \\
24  &   0.31721 & -0.11052 &   1.14 &   49.7 &   17.83 $\pm$   3.29 & 1.96 & 1.11 & -24.4 &    \\
25  &   0.31270 & -0.12153 &   4.42 &   49.7 &   83.72 $\pm$   3.14 & 1.61 & 1.12 & -12.1 &    \\
26  &   0.31129 & -0.07745 &   1.97 &   49.7 &   23.52 $\pm$   3.73 & 2.35 & 1.22 & -26.0 &    \\
27  &   0.30683 & -0.09038 &   4.17 &   49.7 &   81.35 $\pm$   3.76 & 1.70 & 1.03 & -13.3 &    \\
28  &   0.29177 & -0.08839 &   2.17 &  -82.9 &   31.23 $\pm$   3.35 & 1.90 & 1.25 &   7.5 &    \\
29  &   0.28870 &  0.03819 &  13.81 &    0.0 &   29.12 $\pm$   5.41 & 6.00 & 2.70 &-191.0 &    \\
30  &   0.28826 &  0.03358 &   3.90 &    0.0 &   53.23 $\pm$  10.51 & 2.88 & 0.87 &  -1.4 &    \\
31  &   0.28790 &  0.02572 &   9.46 &    0.0 &   28.58 $\pm$   5.07 & 6.00 & 1.22 &   1.2 &    \\
32  &   0.28778 &  0.03178 &   6.13 &    0.0 &   95.57 $\pm$   8.96 & 3.38 & 1.00 &  -1.5 &    \\
33  &   0.28282 &  0.00865 &   3.58 &  116.0 &   35.65 $\pm$   3.94 & 3.01 & 1.14 &  -9.1 &    \\
34  &   0.28200 &  0.04905 &   1.37 &  116.0 &   17.40 $\pm$   2.48 & 2.99 & 0.90 &  -0.0 &    \\
35  &   0.27847 &  0.05379 &   5.77 &    0.0 &   44.92 $\pm$   6.91 & 4.18 & 1.05 &   0.6 &    \\
36  &   0.27824 &  0.03784 &   7.63 &    0.0 &   91.57 $\pm$  12.65 & 2.79 & 1.02 &  -4.2 &    \\
37  &   0.26411 &  0.03025 &  35.72 &    0.0 &  529.92 $\pm$  18.02 & 2.53 & 0.91 &  -1.6 &    \\
38  &   0.26334 &  0.03499 &  17.44 &   33.2 &  203.29 $\pm$  25.65 & 2.64 & 1.11 &  -3.8 &    \\
39  &   0.26076 & -0.06171 &   5.56 &   82.9 &   43.47 $\pm$   4.59 & 2.73 & 1.60 &  -2.8 &    \\
40  &   0.25660 &  0.03957 &   9.54 &   16.6 &   76.30 $\pm$   9.94 & 3.68 & 1.16 &   4.2 &    \\
41  &   0.25344 &  0.02118 &  21.70 &   33.2 &  260.14 $\pm$  26.66 & 2.74 & 1.04 &  -3.7 &    \\
42  &   0.25338 &  0.01865 &  20.38 &   33.2 &  161.11 $\pm$  26.99 & 3.79 & 1.14 &  -8.4 &    \\
43  &   0.25160 &  0.01312 &  20.04 &   16.6 &  165.75 $\pm$  11.84 & 3.33 & 1.24 &  -1.3 &    \\
44  &   0.24827 &  0.00324 &  15.15 &   16.6 &  127.00 $\pm$  20.79 & 3.26 & 1.25 &-191.3 &    \\
45  &   0.24654 &  0.00297 &   6.82 &   33.2 &  274.56 $\pm$  39.00 & 3.77 & 1.25 &   5.1 &    \\
46  &   0.24592 & -0.05180 &  37.88 &   66.3 &   99.69 $\pm$   4.09 & 2.29 & 1.02 &  -4.0 &    \\
47  &   0.24153 & -0.06674 &   1.36 &   66.3 &   22.41 $\pm$   3.11 & 1.96 & 1.06 & -12.4 &    \\
48  &   0.23342 & -0.00047 &  32.02 &   33.2 &  497.48 $\pm$  56.52 & 2.29 & 0.96 &  -2.1 &    \\
49  &   0.23106 & -0.00998 &  24.26 &   33.2 &   88.54 $\pm$  14.88 & 6.00 & 1.56 &   4.5 &    \\
50  &   0.22401 &  0.01991 &   2.53 &   16.6 &   28.20 $\pm$   4.76 & 2.81 & 1.09 &  -1.5 &    \\
51  &   0.21364 & -0.01632 &   5.03 &   82.9 &   19.89 $\pm$   3.37 & 4.91 & 1.76 &-174.9 &    \\
52  &   0.21266 & -0.03738 &   2.03 &   33.2 &   25.03 $\pm$   3.44 & 2.50 & 1.11 &  -6.9 &    \\
53  &   0.20839 &  0.00144 &   0.61 &   49.7 &   12.89 $\pm$   2.58 & 1.93 & 0.84 &  -9.3 &    \\
54  &   0.20507 & -0.02072 &   6.15 &   33.2 &  101.53 $\pm$   4.52 & 2.25 & 0.92 &  -4.5 &    \\
55  &   0.20259 &  0.04108 &   1.17 &   82.9 &   13.55 $\pm$   2.64 & 2.82 & 1.05 &  -4.1 &    \\
56  &   0.20142 & -0.01971 &   1.48 &   82.9 &   84.07 $\pm$   4.40 & 2.61 & 1.11 &  -0.7 &    \\
57  &   0.20089 &  0.02738 &   1.49 &   49.7 &   16.36 $\pm$   3.00 & 2.69 & 1.16 &  -5.7 &    \\
58  &   0.20055 & -0.03783 &   7.13 &   33.2 &   20.26 $\pm$   3.67 & 2.35 & 1.06 &   0.7 &    \\
59  &   0.19687 & -0.03728 &  10.58 &   82.9 &   94.06 $\pm$   5.37 & 3.23 & 1.19 &   0.9 &    \\
60  &   0.19597 &  0.00333 &   4.64 &   33.2 &   39.76 $\pm$   3.46 & 3.47 & 1.15 &  -2.6 &    \\
61  &   0.19254 & -0.06985 &   2.82 &   82.9 &   23.88 $\pm$   4.44 & 3.25 & 1.24 &   6.9 &    \\
62  &   0.19253 & -0.05463 &   3.29 &   16.6 &   20.65 $\pm$   3.84 & 4.28 & 1.27 &   4.4 &    \\
63  &   0.18750 & -0.05862 &   1.07 &   82.9 &   22.75 $\pm$   4.50 & 1.84 & 0.87 & -20.5 &    \\
64  &   0.18532 & -0.04157 &   0.96 &   82.9 &   21.43 $\pm$   4.26 & 1.55 & 0.99 &  -0.5 &    \\
65  &   0.18500 &  0.00011 &   2.02 &   49.7 &   24.91 $\pm$   3.38 & 2.56 & 1.08 &   0.3 &    \\
66  &   0.18302 &  0.02401 &   2.67 &   66.3 &   25.61 $\pm$   3.33 & 3.49 & 1.02 &   1.0 &    \\
67  &   0.18011 & -0.06884 &   9.86 &    0.0 &   37.17 $\pm$   3.97 & 6.00 & 1.51 & 198.7 &    \\
68  &   0.17952 & -0.02330 &   2.32 &  -66.3 &   27.16 $\pm$   2.96 & 2.30 & 1.27 &  -6.4 &    \\
69  &   0.16140 & -0.08581 &   3.77 &   82.9 &   49.32 $\pm$   5.46 & 2.07 & 1.26 &  -3.4 &    \\
70  &   0.16114 & -0.05576 &   3.52 &   16.6 &   23.70 $\pm$   3.46 & 3.76 & 1.35 &  10.9 &    \\
71  &   0.16007 &  0.03726 &   3.81 &   82.9 &   30.73 $\pm$   3.33 & 2.99 & 1.36 &   5.4 &    \\
72  &   0.15974 & -0.08534 &   3.66 &   82.9 &   26.39 $\pm$   4.59 & 2.68 & 1.84 & 219.0 &    \\
73  &   0.15677 & -0.09824 &   2.31 &   66.3 &   27.06 $\pm$   5.00 & 2.54 & 1.15 & 231.6 &    \\
74  &   0.15653 & -0.09594 &   1.10 &   66.3 &   24.69 $\pm$   4.82 & 1.50 & 1.01 & -17.4 &    \\
75  &   0.15609 & -0.01274 &   1.48 &   82.9 &   21.25 $\pm$   2.61 & 2.17 & 1.10 & 261.6 &    \\
76  &   0.15451 &  0.03194 &   5.23 &   99.5 &   14.58 $\pm$   2.33 & 3.14 & 1.17 & 251.8 &    \\
77  &   0.15415 & -0.14875 &   1.57 &  -99.5 &  112.04 $\pm$   4.12 & 2.02 & 0.79 & 225.6 &    \\
78  &   0.15293 & -0.10582 &   2.55 &   66.3 &   29.40 $\pm$   5.04 & 2.45 & 1.21 &  54.9 &    \\
79  &   0.15222 & -0.14471 &   0.43 &  -99.5 &   12.36 $\pm$   2.28 & 1.75 & 0.68 & 228.7 &    \\
80  &   0.15219 &  0.01044 &   1.05 &   82.9 &   24.69 $\pm$   2.66 & 1.91 & 0.76 & 246.0 &    \\
81  &   0.15149 &  0.03119 &   1.97 &   82.9 &   33.26 $\pm$   3.79 & 2.02 & 1.00 & 252.9 &    \\
82  &   0.14727 & -0.09708 &   1.58 &   82.9 &   22.94 $\pm$   3.20 & 2.06 & 1.14 & 234.6 &    \\
83  &   0.14591 &  0.00659 &   0.38 &   49.7 &   42.49 $\pm$   7.24 & 2.29 & 1.25 &  73.0 &    \\
84  &   0.14583 & -0.12922 &   1.50 &   66.3 &   40.17 $\pm$   1.58 & 0.92 & 0.35 &-133.8 &    \\
85  &   0.14560 & -0.12644 &   3.56 &   66.3 &   23.01 $\pm$   3.35 & 2.19 & 1.02 & 223.5 &    \\
86  &   0.14358 &  0.00306 &  35.02 &   66.3 &  294.05 $\pm$  13.22 & 3.60 & 1.13 &  78.1 &    \\
87  &   0.14240 & -0.09352 &   0.19 &  265.2 &    7.41 $\pm$   1.48 & 1.62 & 0.54 & 233.6 &    \\
88  &   0.14191 & -0.04806 &   1.60 &   33.2 &   24.72 $\pm$   3.24 & 2.07 & 1.07 & 238.8 &    \\
89  &   0.14078 & -0.08507 &   1.75 &  -16.6 &   36.02 $\pm$   2.21 & 1.89 & 0.88 & 234.6 &    \\
90  &   0.13691 &  0.03107 &   1.43 &  -16.6 &   16.86 $\pm$   3.29 & 1.95 & 0.86 & 251.6 &    \\
91  &   0.13659 & -0.14041 &   0.83 &   33.2 &   29.08 $\pm$   4.48 & 1.91 & 0.88 & 229.2 &    \\
92  &   0.13648 &  0.00287 &   2.34 &  132.6 &   27.91 $\pm$   3.59 & 2.09 & 1.37 & 252.4 &    \\
93  &   0.13234 &  0.02286 &   2.87 &  -33.2 &   56.73 $\pm$   3.91 & 2.11 & 0.82 &  67.6 &    \\
94  &   0.13061 & -0.12955 &  19.32 &   49.7 &   38.86 $\pm$   4.56 & 6.00 & 2.83 &  32.8 &    \\
95  &   0.12902 &  0.02518 &   0.49 &  -16.6 &  110.21 $\pm$   6.75 & 6.00 & 1.58 & 242.9 &    \\
96  &   0.12883 &  0.02502 &  30.59 &  -16.6 &   59.84 $\pm$   3.25 & 0.75 & 0.37 & -93.3 &    \\
97  &   0.12849 & -0.07990 &  30.90 &   16.6 &   74.86 $\pm$   5.32 & 2.13 & 0.89 & 230.3 &    \\
98  &   0.12762 & -0.09609 &   4.15 &   16.6 &  350.35 $\pm$  10.84 & 2.49 & 1.21 &  40.0 &    \\
99  &   0.12743 & -0.09518 &   8.15 &   16.6 &  113.85 $\pm$  10.35 & 1.62 & 1.51 & 320.3 &    \\
100 &   0.11942 & -0.11160 &  12.28 &   33.2 &  126.28 $\pm$  23.12 & 2.39 & 1.39 & 227.8 &    \\
101 &   0.11876 & -0.03267 &   5.22 &   66.3 &   54.16 $\pm$   9.44 & 2.42 & 1.36 &  58.2 &    \\
102 &   0.11788 & -0.11073 &  68.67 &   33.2 &   87.45 $\pm$  15.63 & 6.00 & 4.47 & 255.2 &    \\
103 &   0.11669 & -0.10904 &   1.33 &   82.9 &   17.11 $\pm$   2.99 & 2.51 & 1.06 & 234.6 &    \\
104 &   0.11254 & -0.01404 &   1.41 &   66.3 &  247.52 $\pm$  21.43 & 2.30 & 1.06 &  61.8 &    \\
105 &   0.11200 & -0.04660 &  17.67 &  -66.3 &   25.94 $\pm$   2.99 & 1.85 & 1.00 & 237.9 &    \\
106 &   0.11066 & -0.00078 &  40.61 &   66.3 &  700.63 $\pm$  38.37 & 2.20 & 0.90 & 246.6 &    \\
107 &   0.11045 &  0.00267 &  62.20 &   49.7 &  201.28 $\pm$   6.09 & 6.00 & 4.39 & 205.2 &    \\
108 &   0.10984 & -0.08124 & 155.22 &   66.3 &  533.96 $\pm$  69.07 & 3.49 & 1.14 & 236.6 &    \\
109 &   0.10925 & -0.09108 &  43.97 &   49.7 &  241.20 $\pm$  43.31 & 3.44 & 1.81 &  71.1 &    \\
110 &   0.10797 & -0.08989 &  33.18 &   49.7 &  376.04 $\pm$  63.78 & 2.20 & 1.37 &  56.0 &    \\
111 &   0.10729 & -0.08458 & 161.85 &   49.7 & 1510.15 $\pm$  19.73 & 2.56 & 1.43 &  55.2 &    \\
112 &   0.10535 & -0.00312 &  12.79 &   33.2 &  179.19 $\pm$  14.86 & 2.30 & 1.06 &  64.2 &    \\
113 &   0.10402 & -0.08034 & 197.63 &   49.7 & 1783.81 $\pm$ 127.98 & 2.38 & 1.59 &  68.8 &    \\
114 &   0.10344 & -0.08196 & 104.14 &   49.7 & 1004.45 $\pm$ 149.02 & 2.27 & 1.56 &  41.9 &    \\
115 &   0.10297 & -0.00477 &   1.85 &   49.7 &  198.27 $\pm$  34.46 & 2.06 & 1.50 & 247.5 &    \\
116 &   0.10273 &  0.05847 &  17.94 &   99.5 &   27.56 $\pm$   4.53 & 2.52 & 0.91 &  77.5 &    \\
117 &   0.10087 &  0.05790 &   3.54 &   99.5 &   50.85 $\pm$   4.92 & 2.45 & 0.97 &  72.8 &    \\
118 &   0.10077 & -0.00734 & 166.60 &   66.3 &  640.39 $\pm$  38.49 & 5.66 & 1.57 & 255.6 &    \\
119 &   0.10071 & -0.00614 &  46.34 &   66.3 &  736.44 $\pm$  49.00 & 1.99 & 1.08 &  64.4 &    \\
120 &   0.09995 & -0.00756 &  20.52 &   66.3 &  429.80 $\pm$  31.75 & 1.23 & 0.40 &  62.9 &    \\
121 &   0.09981 & -0.08000 &   6.19 &   49.7 &  442.89 $\pm$  51.45 & 1.84 & 0.86 & 234.1 &    \\
122 &   0.09785 & -0.02001 &  15.48 &   49.7 &  205.16 $\pm$  13.72 & 2.13 & 1.21 &  64.3 &    \\
123 &   0.09694 & -0.06808 &   6.04 &   33.2 &   47.17 $\pm$   9.21 & 2.34 & 1.87 & 223.1 &    \\
124 &   0.09613 & -0.07940 &  53.96 &   66.3 &  258.23 $\pm$  27.04 & 3.09 & 2.31 &  78.9 &    \\
125 &   0.09154 & -0.11098 &   1.46 &   82.9 &   19.08 $\pm$   2.96 & 2.12 & 1.23 & 235.3 &    \\
126 &   0.08815 & -0.09775 &   2.98 &   16.6 &   48.30 $\pm$   5.83 & 2.15 & 0.98 & 231.0 &    \\
127 &   0.08575 & -0.03094 &   5.55 &   49.7 &  101.63 $\pm$   6.65 & 2.12 & 0.88 &  67.4 &    \\
128 &   0.08339 & -0.05188 &   3.48 &   33.2 &   56.37 $\pm$   9.69 & 1.97 & 1.07 & 232.9 &    \\
129 &   0.08184 & -0.14598 &   2.25 & -116.0 &   30.14 $\pm$   2.37 & 2.38 & 1.07 & 226.0 &    \\
130 &   0.07616 & -0.14549 &   4.14 &   16.6 &   44.51 $\pm$   5.50 & 2.54 & 1.25 & 233.2 &    \\
131 &   0.07527 & -0.07269 &   8.31 &   66.3 &  215.12 $\pm$  19.99 & 2.58 & 1.31 & 241.4 &    \\
132 &   0.07522 & -0.14753 &  21.29 &   16.6 &   31.64 $\pm$   5.07 & 3.56 & 2.52 &-222.1 &    \\
133 &   0.07394 & -0.11272 &   2.22 &   82.9 &   29.39 $\pm$   2.84 & 1.91 & 1.35 & 229.4 &    \\
134 &   0.07324 & -0.08184 &   6.75 &   49.7 &  168.02 $\pm$  19.80 & 1.83 & 0.75 & 231.3 &    \\
135 &   0.06623 & -0.08346 &  27.60 &   49.7 &  449.88 $\pm$  45.72 & 2.16 & 0.97 & 235.7 &    \\
136 &   0.06567 & -0.03929 &  11.02 &   33.2 &  202.25 $\pm$   7.12 & 2.19 & 0.85 &  59.7 &    \\
137 &   0.06264 & -0.07933 &  41.42 &   49.7 &  424.97 $\pm$  54.89 & 2.19 & 1.52 &  55.7 &    \\
138 &   0.06264 & -0.07933 &  41.40 &   49.7 &  424.83 $\pm$  54.88 & 2.19 & 1.52 &  55.8 &    \\
139 &   0.06220 & -0.06987 &  28.65 &   49.7 &  342.17 $\pm$  31.17 & 2.15 & 1.33 &  59.7 &    \\
140 &   0.05921 & -0.07542 &  42.88 &   49.7 &  417.09 $\pm$  43.40 & 2.28 & 1.54 &  49.5 &    \\
141 &   0.05832 & -0.02368 &  22.13 &   49.7 &  483.74 $\pm$  11.51 & 1.86 & 0.84 & 242.4 &    \\
142 &   0.05747 & -0.07676 & 100.26 &   33.2 & 1482.48 $\pm$  36.40 & 2.20 & 1.05 & 232.2 &    \\
143 &   0.05258 & -0.02065 &   3.69 &   33.2 &   85.68 $\pm$   5.06 & 1.84 & 0.80 & 238.4 &    \\
144 &   0.03621 & -0.00703 &   0.86 &    0.0 &   14.08 $\pm$   2.57 & 2.08 & 1.00 & 233.3 &    \\
145 &   0.02881 &  0.03942 &   0.99 &   66.3 &   14.10 $\pm$   2.74 & 2.47 & 0.97 &  73.4 &    \\
146 &   0.02861 & -0.03854 &   5.99 &   49.7 &   88.81 $\pm$  11.47 & 2.45 & 0.94 & 240.7 &    \\
147 &   0.02850 & -0.07745 &   1.71 &  -16.6 &   29.24 $\pm$   3.30 & 2.06 & 0.97 & 238.6 &    \\
148 &   0.02545 &  0.03683 &   9.25 &   82.9 &  110.31 $\pm$   9.50 & 2.58 & 1.11 &  72.7 &    \\
149 &   0.02288 & -0.08630 &   3.06 &   66.3 &   36.11 $\pm$   5.23 & 2.58 & 1.12 & 231.5 &    \\
150 &   0.02208 &  0.02794 &   5.61 &   82.9 &   69.64 $\pm$   8.60 & 2.20 & 1.25 &  75.4 &    \\
151 &   0.02084 &  0.03667 &  91.15 &   99.5 &   63.09 $\pm$   6.50 & 2.38 & 1.16 & 254.0 &    \\
152 &   0.02055 & -0.05013 &   5.10 &    0.0 &  109.47 $\pm$  11.74 & 6.00 & 4.74 & 353.3 &    \\
153 &   0.01534 &  0.03769 &   8.91 &   82.9 &  146.81 $\pm$  12.86 & 2.16 & 0.96 &  68.7 &    \\
154 &   0.01145 & -0.00382 &   1.09 &   33.2 &   21.80 $\pm$   2.47 & 2.09 & 0.82 & 240.5 &    \\
155 &   0.01023 & -0.01611 &   4.80 &    0.0 &   57.24 $\pm$   5.17 & 2.63 & 1.09 &  56.1 &    \\
156 & 359.99392 & -0.08895 &  15.48 &   16.6 &  196.57 $\pm$  13.58 & 2.36 & 1.14 &  58.7 &    \\
157 & 359.99307 & -0.08222 &  11.83 &   33.2 &  157.88 $\pm$  27.20 & 2.15 & 1.19 & 230.9 &    \\
158 & 359.99125 & -0.04605 &  24.35 &   82.9 &  200.40 $\pm$  13.24 & 2.15 & 1.93 &  45.5 &    \\
159 & 359.98948 & -0.06961 &  36.55 &   49.7 & 4640.42 $\pm$   9.75 & 2.15 & 1.18 &  54.9 &    \\
160 & 359.98945 & -0.05699 & 344.67 &   66.3 &  334.04 $\pm$  41.22 & 1.92 & 0.95 & 235.9 & SPF 2010   \\
161 & 359.98874 & -0.08297 &  17.84 &   16.6 &  353.04 $\pm$  22.55 & 2.60 & 1.36 &  47.1 &   \\
162 & 359.98733 & -0.09945 &   0.62 & -132.6 &   14.30 $\pm$   2.30 & 1.89 & 0.78 & 224.4 &    \\
163 & 359.98725 & -0.04716 &   7.12 &   66.3 &  133.69 $\pm$  19.12 & 1.82 & 1.00 & 237.7 &    \\
164 & 359.98442 & -0.06142 & 468.31 &   49.7 & 7260.27 $\pm$  56.31 & 2.16 & 1.02 & 242.8 &  SPF 2010  \\
165 & 359.98313 & -0.06307 & 243.67 &   49.7 & 3847.74 $\pm$ 256.27 & 2.23 & 0.97 & 236.4 & SPF 2010   \\
166 & 359.98229 & -0.04431 &   6.66 &    0.0 &   68.40 $\pm$   8.44 & 2.60 & 1.28 &  69.3 &    \\
167 & 359.97477 & -0.07051 & 238.62 &   49.7 & 4403.94 $\pm$ 198.13 & 1.99 & 0.93 & 234.3 &  SPF 2010  \\
168 & 359.96946 & -0.07529 & 270.76 &   49.7 & 3086.45 $\pm$ 127.20 & 2.27 & 1.32 & 243.3 &  SPF 2010  \\
169 & 359.95024 & -0.05775 &   6.46 &   16.6 &   94.22 $\pm$   7.27 & 2.17 & 1.08 & 235.4 &  SPF 2010  \\
170 & 359.94841 &  0.01754 &   4.89 &   66.3 &   20.95 $\pm$   3.60 & 1.99 & 1.21 &  67.1 &    \\
171 & 359.94789 & -0.07981 &   1.48 &   33.2 &   73.30 $\pm$  11.92 & 1.98 & 1.15 &  54.6 &    \\
172 & 359.94634 &  0.09725 &   1.51 &   82.9 &   25.52 $\pm$   4.66 & 1.81 & 1.12 &  72.7 &    \\
173 & 359.94603 & -0.08819 &   5.47 &   33.2 &   49.17 $\pm$   7.09 & 3.22 & 1.18 & 241.7 &    \\
174 & 359.94057 & -0.09195 &   3.35 &   66.3 &   61.44 $\pm$   3.28 & 2.07 & 0.90 & 237.2 &    \\
175 & 359.93894 & -0.12680 &   1.51 &  -99.5 &   17.42 $\pm$   2.43 & 2.43 & 1.22 & 231.4 &    \\
176 & 359.93748 &  0.02629 &  11.31 &   66.3 &   44.72 $\pm$   5.65 & 4.57 & 1.89 & 105.7 &    \\
177 & 359.93498 & -0.12659 &   1.67 &  -99.5 &   14.26 $\pm$   2.56 & 2.48 & 1.61 &  19.4 &    \\
178 & 359.93353 &  0.02722 &   3.71 &   66.3 &   29.82 $\pm$   5.68 & 1.81 & 0.79 &  71.7 &    \\
179 & 359.93280 & -0.05830 &   1.25 &   49.7 &   22.49 $\pm$   4.22 & 4.34 & 1.30 & 228.5 &    \\
180 & 359.93145 & -0.08252 &  14.43 &   33.2 &  106.18 $\pm$   9.84 & 2.92 & 1.59 &  41.8 &    \\
181 & 359.92907 & -0.10853 &   0.87 & -132.6 &   12.29 $\pm$   2.18 & 2.10 & 1.15 &  60.3 &    \\
182 & 359.91572 & -0.13387 &   1.98 & -149.2 &   12.37 $\pm$   2.01 & 4.63 & 1.18 & 218.8 &    \\
183 & 359.91330 & -0.03368 &   1.68 &    0.0 &   33.61 $\pm$   2.66 & 1.94 & 0.88 & 246.4 &    \\
184 & 359.91238 & -0.06231 &  13.31 &   16.6 &  171.46 $\pm$  20.79 & 2.41 & 1.10 &  62.9 &    \\
185 & 359.91212 & -0.05326 &   8.94 &   16.6 &   76.44 $\pm$  11.29 & 2.48 & 1.61 & 238.1 &    \\
186 & 359.91146 & -0.13058 &   4.44 &   66.3 &   43.89 $\pm$   2.87 & 2.52 & 1.37 &  54.6 &    \\
187 & 359.90982 & -0.12498 &   2.12 & -149.2 &   36.64 $\pm$   2.39 & 2.04 & 0.97 & 232.8 &    \\
188 & 359.90637 & -0.10605 &   3.74 & -116.0 &   46.75 $\pm$   2.36 & 2.60 & 1.05 &  53.3 &    \\
189 & 359.90143 & -0.06112 &  33.62 &   16.6 &  193.26 $\pm$  27.05 & 6.00 & 1.69 & 248.8 &    \\
190 & 359.90064 & -0.07022 &  57.37 &   16.6 &  293.48 $\pm$  43.29 & 2.43 & 1.61 & 257.3 &    \\
191 & 359.90017 & -0.06338 &   6.42 &    0.0 &   65.62 $\pm$   9.91 & 2.32 & 1.44 & 244.8 &    \\
192 & 359.89990 &  0.01417 &  64.23 &   82.9 &   94.25 $\pm$   4.99 & 6.00 & 3.88 & 177.8 &    \\
193 & 359.89776 &  0.01630 &   4.44 &   82.9 &   81.03 $\pm$   8.06 & 2.08 & 0.90 &  68.0 &    \\
194 & 359.89603 &  0.01909 &   5.45 &   66.3 &   36.40 $\pm$   6.43 & 3.14 & 1.63 & 266.2 &    \\
195 & 359.89544 &  0.02174 &   7.71 &   99.5 &   16.47 $\pm$   3.18 & 2.27 & 1.14 &  69.7 &    \\
196 & 359.89540 & -0.09942 &   1.25 &   16.6 &   78.15 $\pm$   7.19 & 2.34 & 1.44 &  68.4 &    \\
197 & 359.89370 & -0.06916 &  87.53 &   16.6 &  466.54 $\pm$  65.25 & 3.56 & 1.80 & 260.7 &    \\
198 & 359.89341 & -0.09847 &   4.85 &   33.2 &   25.76 $\pm$   4.22 & 3.99 & 1.61 &  76.2 &    \\
199 & 359.89247 &  0.02356 &  23.71 &   82.9 &   20.52 $\pm$   4.00 & 1.95 & 0.82 & 252.4 &    \\
200 & 359.89216 & -0.07664 &   8.91 &   16.6 &  322.48 $\pm$  34.68 & 2.73 & 0.92 &  56.4 &    \\
201 & 359.89188 & -0.06567 &  32.31 &   16.6 & 1128.26 $\pm$  50.52 & 1.63 & 0.60 & 246.6 &    \\
202 & 359.89173 & -0.06563 & 121.66 &   16.6 &  685.48 $\pm$  53.46 & 4.21 & 1.44 &  82.0 &    \\
203 & 359.89166 & -0.06671 &   0.96 &   33.2 &  237.82 $\pm$  41.04 & 1.60 & 0.80 &  60.9 &    \\
204 & 359.89062 & -0.05378 &   2.77 &    0.0 &   48.68 $\pm$   4.57 & 2.16 & 0.90 & 240.8 &    \\
205 & 359.88871 & -0.06663 &  15.40 &   33.2 &  219.32 $\pm$  30.02 & 2.05 & 1.17 &  63.6 &    \\
206 & 359.87897 & -0.08370 &  67.67 &   16.6 &   64.21 $\pm$   8.61 & 6.00 & 6.00 &-308.3 &    \\
207 & 359.87722 & -0.09400 &   2.96 &   33.2 &   36.71 $\pm$   5.75 & 2.15 & 1.28 & 232.5 &    \\
208 & 359.87322 & -0.08219 &   6.97 &    0.0 &   74.28 $\pm$   9.96 & 2.94 & 1.09 &  58.1 &    \\
209 & 359.87316 & -0.06802 &   5.16 &    0.0 &  113.50 $\pm$  12.33 & 1.94 & 0.80 & 238.7 &    \\
210 & 359.87288 & -0.04584 &   3.45 &   16.6 &   38.97 $\pm$   4.78 & 2.03 & 1.49 & 233.3 &    \\
211 & 359.86896 & -0.05139 &   5.55 &    0.0 &   87.10 $\pm$   6.00 & 2.20 & 0.99 &  59.3 &    \\
212 & 359.86736 & -0.10064 &   0.77 &   16.6 &   34.22 $\pm$   6.60 & 3.52 & 1.07 &  48.8 &    \\
213 & 359.86727 & -0.05074 &   3.77 &    0.0 &   47.01 $\pm$   6.83 & 2.04 & 1.15 & 243.4 &    \\
214 & 359.86726 & -0.10397 &   3.23 &   16.6 &   23.94 $\pm$   4.48 & 1.50 & 0.73 & 228.7 &    \\
215 & 359.86319 & -0.08024 &   9.43 &    0.0 &   85.19 $\pm$  16.18 & 1.97 & 1.92 &  57.6 &    \\
216 & 359.85787 & -0.08236 &   8.12 &    0.0 &   88.36 $\pm$  13.14 & 1.95 & 1.61 &  66.5 &    \\
217 & 359.85638 &  0.02244 &   0.86 &   66.3 &   25.21 $\pm$   2.50 & 2.77 & 1.25 & 253.8 &    \\
218 & 359.85585 & -0.07471 &   2.56 &   66.3 &   15.06 $\pm$   2.25 & 1.99 & 0.98 & 238.0 &    \\
219 & 359.85547 & -0.08292 &   9.34 &   16.6 &  101.81 $\pm$  18.73 & 2.59 & 1.21 &  63.5 &    \\
220 & 359.85228 & -0.08624 & 149.18 &    0.0 &  408.28 $\pm$   0.49 & 6.00 & 2.08 &  80.6 &    \\
221 & 359.85105 & -0.07944 &  15.12 &    0.0 &  161.34 $\pm$  31.82 & 2.56 & 1.25 &  61.5 &    \\
222 & 359.84987 & -0.08684 &  22.38 &    0.0 &  214.02 $\pm$  32.45 & 2.48 & 1.44 &  52.8 &    \\
223 & 359.84749 & -0.06893 &  10.98 &    0.0 &   88.10 $\pm$  11.41 & 2.93 & 2.12 & -86.6 &    \\
224 & 359.84692 & -0.09060 &  16.02 &    0.0 &  194.83 $\pm$  19.28 & 2.07 & 0.93 & 236.2 &    \\
225 & 359.83829 & -0.06929 &   0.53 &   33.2 &   14.30 $\pm$   2.10 & 1.84 & 0.69 & 242.6 &    \\
226 & 359.83730 & -0.06854 &   4.39 &   16.6 &   89.22 $\pm$   3.52 & 2.00 & 0.84 & 239.1 &    \\
227 & 359.82592 &  0.02142 &  37.72 &   66.3 &  139.44 $\pm$  12.17 & 6.00 & 1.54 & 249.3 &    \\
228 & 359.82202 & -0.07485 &   1.37 &   33.2 &   41.98 $\pm$   3.54 & 2.39 & 1.11 & 233.6 &    \\
229 & 359.82190 & -0.10264 &   3.26 &    0.0 &   20.19 $\pm$   3.10 & 2.00 & 1.16 & 233.0 &    \\
230 & 359.82047 &  0.01241 &   3.54 &   66.3 &   39.66 $\pm$   4.20 & 2.53 & 1.02 &  70.2 &    \\
231 & 359.82038 & -0.11362 &   3.00 &    0.0 &   21.64 $\pm$   2.12 & 2.94 & 1.90 & 211.6 &    \\
232 & 359.81900 & -0.08750 &   0.50 &   16.6 &   13.02 $\pm$   2.47 & 1.92 & 0.68 & 226.9 &    \\
233 & 359.81586 &  0.01507 &   6.22 &   66.3 &   20.33 $\pm$   3.41 & 5.94 & 1.76 &  42.5 &    \\
234 & 359.81546 & -0.08071 &   2.52 &  -16.6 &   42.04 $\pm$   2.88 & 2.47 & 0.83 &  59.1 &    \\
235 & 359.81315 & -0.06568 &   2.53 &   33.2 &   15.99 $\pm$   2.47 & 3.51 & 1.54 & 220.9 &    \\
236 & 359.78621 & -0.07778 &   0.61 &   33.2 &   12.47 $\pm$   2.24 & 2.07 & 0.81 &  61.1 &    \\
237 & 359.78515 & -0.07612 &   0.93 &   16.6 &   19.36 $\pm$   2.07 & 2.01 & 0.82 & 235.2 &    \\
238 & 359.77925 & -0.14674 &   1.28 &   16.6 &   14.88 $\pm$   2.46 & 3.22 & 0.91 &  44.6 &    \\
239 & 359.77882 & -0.08390 &   4.41 &  -33.2 &   40.03 $\pm$   2.74 & 3.58 & 1.05 & 236.8 &    \\
240 & 359.76561 &  0.00464 &   0.91 &   66.3 &   15.82 $\pm$   2.34 & 1.96 & 1.00 & 245.1 &    \\
241 & 359.75855 & -0.10280 &   0.56 &   66.3 &   16.14 $\pm$   2.12 & 1.66 & 0.72 & 238.1 &    \\
242 & 359.74707 & -0.10923 &   3.12 &  -16.6 &   57.76 $\pm$   2.20 & 2.17 & 0.85 & 234.6 &    \\
243 & 359.74042 &  0.02272 &   1.13 & -116.0 &   14.10 $\pm$   2.46 & 2.34 & 1.17 & 247.6 &    \\
244 & 359.73899 &  0.01358 &   1.40 &  -82.9 &   10.34 $\pm$   1.95 & 2.83 & 1.63 & 242.1 &    \\
245 & 359.73609 & -0.07928 &   4.24 &   99.5 &   44.91 $\pm$   2.45 & 2.60 & 1.24 & 223.0 &    \\
246 & 359.73059 & -0.16664 &   3.99 &  -33.2 &   74.02 $\pm$   8.66 & 1.96 & 0.94 & 224.6 &    \\
247 & 359.73010 & -0.04932 &   4.28 &   16.6 &   18.96 $\pm$   3.52 & 1.86 & 0.96 & 246.5 &    \\
248 & 359.73005 & -0.06570 &   0.99 &  -66.3 &   14.77 $\pm$   2.02 & 6.00 & 1.65 &  65.5 &    \\
249 & 359.72813 & -0.05850 &   3.36 &  -49.7 &   19.13 $\pm$   3.21 & 3.00 & 2.00 & 234.1 &    \\
250 & 359.72710 &  0.02165 &   3.31 &   66.3 &   31.65 $\pm$   2.27 & 2.98 & 1.20 & 250.3 &    \\
251 & 359.72605 & -0.05994 &   4.47 &  -66.3 &   41.08 $\pm$   2.82 & 2.46 & 1.51 & 232.7 &    \\
252 & 359.72604 &  0.01086 &   0.58 &   66.3 &   12.93 $\pm$   1.96 & 1.80 & 0.85 & 248.8 &    \\
253 & 359.72377 & -0.05948 &   1.24 &   16.6 &   22.02 $\pm$   3.75 & 1.86 & 1.03 &  56.7 &    \\
254 & 359.71997 & -0.04707 &   0.23 &  165.8 &    7.68 $\pm$   1.46 & 1.41 & 0.71 & 239.5 &    \\
255 & 359.71976 & -0.00115 &   9.43 &  -66.3 &   69.86 $\pm$   3.07 & 3.27 & 1.41 &  70.7 &    \\
256 & 359.71863 & -0.05738 &   9.27 &   16.6 &  184.25 $\pm$   3.93 & 1.93 & 0.89 & 240.6 &    \\
257 & 359.71749 & -0.11269 &   2.31 &   49.7 &   38.24 $\pm$   2.11 & 2.17 & 0.95 & 231.0 &    \\
258 & 359.70932 &  0.04683 &   1.91 &  -33.2 &   13.99 $\pm$   2.17 & 2.78 & 1.68 & 241.6 &    \\
259 & 359.70734 & -0.07544 &   3.79 &   16.6 &   29.54 $\pm$   4.12 & 2.53 & 1.73 &  22.5 &    \\
260 & 359.70725 & -0.07365 &   6.22 &  -16.6 &   25.64 $\pm$   3.94 & 3.65 & 2.27 & -68.0 &    \\
261 & 359.70077 &  0.04532 &   0.36 &  -66.3 &    7.77 $\pm$   1.46 & 1.87 & 0.85 & 250.0 &    \\
262 & 359.70041 & -0.06189 &  24.31 &   16.6 &  485.33 $\pm$  12.17 & 1.88 & 0.91 & 236.9 &    \\
263 & 359.70034 & -0.06017 &  16.39 &  -16.6 &   83.65 $\pm$   7.42 & 3.78 & 1.77 & -79.6 &    \\
264 & 359.69811 &  0.04914 &   0.58 &  -66.3 &   12.55 $\pm$   1.74 & 2.02 & 0.78 & 253.6 &    \\
265 & 359.69535 & -0.06408 &  91.21 &    0.0 &  199.05 $\pm$  19.10 & 5.36 & 2.92 &  40.7 &    \\
266 & 359.69505 & -0.00900 &   1.40 &  -33.2 &   17.23 $\pm$   2.99 & 2.27 & 1.22 &  66.6 &    \\
267 & 359.69393 & -0.03167 &   0.99 &  -33.2 &   10.07 $\pm$   1.85 & 2.62 & 1.28 &  43.6 &    \\
268 & 359.69368 & -0.01172 &   1.57 &  -33.2 &   24.77 $\pm$   2.60 & 2.14 & 1.01 &  63.6 &    \\
269 & 359.69215 & -0.06432 &   1.61 &   16.6 &   33.73 $\pm$   5.87 & 1.81 & 0.90 & 233.5 &    \\
270 & 359.69072 & -0.06757 &   1.13 &   49.7 &   16.03 $\pm$   2.77 & 2.13 & 1.13 & 236.5 &    \\
271 & 359.69037 & -0.06198 &   2.13 &   66.3 &   13.26 $\pm$   2.25 & 2.92 & 1.88 &-105.4 &    \\
272 & 359.68993 &  0.02148 &   0.66 &  182.3 &    8.22 $\pm$   1.58 & 2.25 & 1.22 & 244.4 &    \\
273 & 359.68591 & -0.08802 &   2.36 &   16.6 &   29.54 $\pm$   4.64 & 2.07 & 1.32 &  50.3 &    \\
274 & 359.68433 & -0.08619 &   1.98 &   16.6 &   29.68 $\pm$   4.47 & 2.13 & 1.07 & 231.6 &    \\
275 & 359.67418 & -0.08344 &   1.71 &   16.6 &   16.55 $\pm$   2.97 & 2.73 & 1.29 &  68.3 &    \\
276 & 359.67309 & -0.11099 &   2.77 &  -33.2 &   27.40 $\pm$   4.21 & 2.20 & 1.57 & 243.7 &    \\
277 & 359.67181 & -0.11227 &   9.25 &  -49.7 &   48.11 $\pm$   5.24 & 4.35 & 1.51 & 235.2 &    \\
278 & 359.67104 & -0.01738 &   4.10 &  -49.7 &   29.03 $\pm$   2.72 & 3.15 & 1.53 &  51.9 &    \\
279 & 359.66701 & -0.12689 &   3.45 &   82.9 &    9.06 $\pm$   1.78 & 1.60 & 0.50 & 236.0 &    \\
280 & 359.66696 & -0.12769 &   0.21 &   82.9 &   10.57 $\pm$   1.77 & 6.00 & 1.86 &   3.8 &    \\
281 & 359.66485 & -0.00561 &   2.87 &  -66.3 &   26.70 $\pm$   2.73 & 2.62 & 1.40 & 231.4 &    \\
282 & 359.66334 & -0.00443 &   3.35 &  -66.3 &   13.43 $\pm$   2.09 & 3.24 & 2.06 &  95.4 &    \\
283 & 359.66298 & -0.10114 &   2.62 &   33.2 &   46.97 $\pm$   2.88 & 2.30 & 1.06 & 229.8 &    \\
284 & 359.66214 & -0.09612 &   6.43 &   33.2 &   14.26 $\pm$   1.76 & 5.66 & 2.72 &  73.0 &    \\
285 & 359.65748 & -0.04938 &   0.25 &    0.0 &    7.95 $\pm$   1.51 & 1.41 & 0.75 & 241.2 &    \\
286 & 359.65518 & -0.11404 &   3.51 &  -33.2 &   44.25 $\pm$   3.51 & 2.28 & 1.19 & 232.2 &    \\
287 & 359.65331 & -0.11655 &   3.79 &   16.6 &   19.83 $\pm$   2.40 & 3.47 & 1.88 &  20.7 &    \\
288 & 359.65286 & -0.11471 &   2.10 &  -16.6 &   31.73 $\pm$   3.05 & 2.35 & 0.96 & 227.2 &    \\
289 & 359.65240 & -0.07259 &   0.64 &   49.7 &   12.04 $\pm$   2.04 & 2.00 & 0.91 & 235.2 &    \\
290 & 359.64844 & -0.07564 &   0.82 &   66.3 &   13.57 $\pm$   2.01 & 2.26 & 0.91 & 234.4 &    \\
291 & 359.64410 & -0.11235 &   2.34 &  -33.2 &   14.27 $\pm$   1.87 & 3.62 & 1.55 &  78.6 &    \\
292 & 359.63871 & -0.11517 &   0.70 &  -99.5 &   13.09 $\pm$   1.52 & 2.09 & 0.88 & 232.0 &    \\
293 & 359.63413 & -0.00388 &   0.85 &   33.2 &   11.17 $\pm$   2.08 & 2.69 & 0.97 &  69.2 &    \\
294 & 359.63116 & -0.10315 &   2.10 &  -33.2 &   37.20 $\pm$   2.01 & 2.05 & 0.94 & 229.2 &    \\
295 & 359.63046 & -0.12960 &  19.33 &  -33.2 &   42.97 $\pm$   3.54 & 6.00 & 2.56 &  64.7 &    \\
296 & 359.62961 & -0.11936 &   1.77 &    0.0 &   30.20 $\pm$   2.30 & 2.08 & 0.96 & 228.5 &    \\
297 & 359.62697 & -0.00252 &   0.50 &   49.7 &   11.88 $\pm$   1.70 & 1.85 & 0.78 & 241.6 &    \\
298 & 359.62601 & -0.01054 &   3.28 &    0.0 &   13.32 $\pm$   2.38 & 3.20 & 2.63 &   8.3 &    \\
299 & 359.62237 & -0.02023 &   0.87 &   16.6 &   14.53 $\pm$   2.87 & 2.60 & 0.79 &  72.6 &    \\
300 & 359.61799 &  0.00161 &   0.88 &  -82.9 &   15.81 $\pm$   1.71 & 1.89 & 1.01 &  68.1 &    \\
301 & 359.61562 & -0.12052 &   3.56 &  -16.6 &   41.22 $\pm$   2.54 & 2.73 & 1.08 &  51.7 &    \\
302 & 359.61293 & -0.10059 &   0.64 & -165.8 &   10.24 $\pm$   1.92 & 1.98 & 1.07 & 224.2 &    \\
303 & 359.61182 & -0.14008 &   1.22 &   33.2 &   15.89 $\pm$   3.05 & 2.22 & 1.18 & 217.4 &    \\
304 & 359.61137 & -0.15415 &  11.32 &   49.7 &   30.14 $\pm$   5.69 & 4.63 & 2.77 &-172.5 &    \\
305 & 359.61073 & -0.13502 &   4.14 &   16.6 &   35.20 $\pm$   3.71 & 2.89 & 1.39 & 224.4 &    \\
306 & 359.60968 & -0.12697 &  18.78 &   16.6 &   69.86 $\pm$   4.38 & 4.10 & 2.24 & 235.8 &    \\
307 & 359.60539 & -0.12674 &   5.37 &   16.6 &   36.27 $\pm$   4.76 & 3.14 & 1.61 & 236.2 &    \\
308 & 359.60103 & -0.13530 &   2.05 & -116.0 &   29.30 $\pm$   2.68 & 2.21 & 1.08 & 223.4 &    \\
309 & 359.59932 & -0.03160 &   0.26 &   66.3 &   10.47 $\pm$   1.94 & 1.65 & 0.51 & 249.4 &    \\
310 & 359.59745 & -0.10733 &   0.93 &  -33.2 &   12.19 $\pm$   1.58 & 2.23 & 1.17 &  61.4 &    \\
311 & 359.59571 & -0.08651 &   1.44 &   66.3 &   24.90 $\pm$   2.23 & 1.98 & 1.00 & 232.1 &    \\
312 & 359.59532 &  0.01449 &   2.42 &  -82.9 &   14.19 $\pm$   1.77 & 4.05 & 1.44 &  43.5 &    \\
313 & 359.59509 &  0.01421 &   0.33 &  -82.9 &   15.38 $\pm$   1.42 & 1.45 & 0.50 & -97.0 &    \\
314 & 359.59207 & -0.03232 &  33.69 &   16.6 &  661.85 $\pm$   6.70 & 1.89 & 0.92 & 241.7 &    \\
315 & 359.59174 & -0.01153 &   0.41 &  -33.2 &   10.62 $\pm$   1.97 & 1.57 & 0.83 &  65.7 &    \\
316 & 359.59090 &  0.01987 &   0.79 &   49.7 &   13.68 $\pm$   2.17 & 3.62 & 1.17 &  70.8 &    \\
317 & 359.59057 & -0.07270 &   1.70 & -182.3 &   10.37 $\pm$   1.75 & 2.29 & 1.14 &  57.9 &    \\
318 & 359.59024 & -0.03875 &   1.57 &  -16.6 &   36.16 $\pm$   2.19 & 1.76 & 0.84 & 241.8 &    \\
319 & 359.58603 & -0.03510 &   3.23 &    0.0 &   39.98 $\pm$   3.98 & 1.96 & 1.01 & 242.2 &    \\
320 & 359.58557 & -0.12764 &   2.32 &  -33.2 &   48.75 $\pm$   3.84 & 2.22 & 1.02 & 232.1 &    \\
321 & 359.58532 & -0.12387 &   2.00 &  -33.2 &   25.16 $\pm$   2.94 & 2.19 & 1.24 & 235.3 &    \\
322 & 359.58314 & -0.12013 &   5.98 &   49.7 &   66.68 $\pm$   2.81 & 2.27 & 1.35 & 230.9 &    \\
323 & 359.58273 & -0.02309 &   5.41 &  -33.2 &  113.19 $\pm$   3.00 & 1.92 & 0.85 & 242.2 &    \\
324 & 359.58159 & -0.02835 &   1.26 &    0.0 &   18.99 $\pm$   3.25 & 2.00 & 1.13 &  64.7 &    \\
325 & 359.57500 & -0.03812 &   1.59 &   16.6 &   27.08 $\pm$   3.19 & 2.00 & 1.00 &  63.0 &    \\
326 & 359.57453 & -0.08034 &   5.50 &  -33.2 &   36.01 $\pm$   2.89 & 3.55 & 1.47 & 221.2 &    \\
327 & 359.57163 & -0.06838 &  10.87 &   16.6 &  203.60 $\pm$   3.63 & 1.96 & 0.93 & 235.1 &    \\
328 & 359.56911 & -0.04780 &   8.74 &   16.6 &   14.62 $\pm$   2.52 & 1.79 & 0.73 & 241.8 &    \\
329 & 359.56875 & -0.06905 &   0.56 &    0.0 &   38.35 $\pm$   5.11 & 4.50 & 1.73 & 254.9 &    \\
330 & 359.56357 & -0.08833 &   2.03 &   16.6 &   34.29 $\pm$   2.97 & 1.96 & 1.03 & 227.4 &    \\
331 & 359.56338 & -0.07097 &   0.73 &    0.0 &   26.92 $\pm$   4.42 & 2.14 & 1.10 &  61.4 &    \\
332 & 359.56268 & -0.14744 &   1.86 &   16.6 &   10.99 $\pm$   2.19 & 1.96 & 1.15 & 223.6 &    \\
333 & 359.56238 &  0.00319 &   4.32 &  -99.5 &   12.61 $\pm$   1.95 & 5.24 & 1.16 &  66.8 &    \\
334 & 359.56205 & -0.07962 &   2.80 &  -16.6 &   48.77 $\pm$   6.82 & 2.11 & 0.93 & 235.8 &    \\
335 & 359.56173 & -0.10442 &   2.24 &    0.0 &   59.33 $\pm$   3.23 & 2.26 & 1.10 &  46.5 &    \\
336 & 359.56124 & -0.06490 &   4.82 &   16.6 &   34.59 $\pm$   4.62 & 3.55 & 1.77 &  47.5 &    \\
337 & 359.56105 & -0.09965 &   6.36 &  -33.2 &   18.80 $\pm$   3.16 & 6.00 & 1.46 & 243.4 &    \\
338 & 359.56024 & -0.07788 &   4.22 &    0.0 &  201.35 $\pm$   6.58 & 2.40 & 1.47 & 242.2 &    \\
339 & 359.55987 & -0.10264 &  20.80 &  -16.6 &   39.73 $\pm$   4.77 & 2.50 & 1.45 &  47.4 &    \\
340 & 359.55948 & -0.07965 &   2.83 &    0.0 &   88.69 $\pm$   8.18 & 2.30 & 1.25 & 236.9 &    \\
341 & 359.55939 & -0.06581 &   7.47 &   16.6 &   54.88 $\pm$   4.80 & 2.28 & 1.36 & 234.5 &    \\
342 & 359.55915 & -0.09138 &   4.98 &  -33.2 &   43.55 $\pm$   4.96 & 2.20 & 1.01 & 229.3 &    \\
343 & 359.55800 & -0.12721 &   1.75 &   49.7 &   15.75 $\pm$   2.61 & 2.52 & 1.51 &  41.9 &    \\
344 & 359.55656 & -0.02985 &   0.75 &   33.2 &   12.25 $\pm$   2.18 & 1.95 & 1.07 &  62.7 &    \\
345 & 359.55541 & -0.08125 &   0.61 &   49.7 &   12.27 $\pm$   2.17 & 1.88 & 0.91 & 236.7 &    \\
346 & 359.55400 & -0.08780 &   2.10 &  -33.2 &   26.30 $\pm$   3.26 & 2.39 & 1.14 & 227.5 &    \\
347 & 359.55224 & -0.10681 &   0.84 &  -16.6 &   14.87 $\pm$   2.97 & 1.57 & 1.23 & 221.6 &    \\
348 & 359.55190 & -0.09791 &   3.44 &    0.0 &   49.21 $\pm$   5.06 & 2.21 & 1.08 & 226.1 &    \\
349 & 359.54964 & -0.02676 &   1.12 &   16.6 &    9.15 $\pm$   1.79 & 2.63 & 1.59 & -80.5 &    \\
350 & 359.54830 & -0.10225 &   1.79 &    0.0 &   23.91 $\pm$   2.92 & 2.37 & 1.08 & 225.7 &    \\
351 & 359.54553 &  0.02807 &   0.63 &  165.8 &    9.69 $\pm$   1.75 & 2.30 & 0.96 &  68.4 &    \\
352 & 359.54356 & -0.03096 &   0.59 &  132.6 &   12.23 $\pm$   1.97 & 1.76 & 0.93 & 240.6 &    \\
353 & 359.54333 & -0.07923 &   3.83 & -149.2 &   20.59 $\pm$   1.67 & 3.69 & 1.72 &  30.8 &    \\
354 & 359.53838 & -0.00289 &   0.42 &  -99.5 &  104.52 $\pm$   0.91 & 1.97 & 0.07 &  92.8 &    \\
355 & 359.53821 & -0.00291 &   8.84 &  -99.5 &  124.24 $\pm$   5.13 & 2.48 & 0.98 & 247.7 &    \\
\enddata
\end{deluxetable}

\end{document}